\documentclass[a4paper,11pt]{article}
\pdfoutput=1 

\usepackage{jheppub} 

\usepackage[T1]{fontenc} 

\usepackage[utf8]{inputenc}
\usepackage{braket}
\usepackage{amsmath}
\usepackage{mathtools}
\usepackage{amssymb}
\usepackage{amsfonts}
\usepackage{graphicx}
\usepackage{bbold}
\usepackage{color}
\usepackage{makecell}
\usepackage{enumitem}
\usepackage{upgreek}
\usepackage{makecell}

\usepackage[normalem]{ulem}
\usepackage{amsmath,amsthm,amssymb}
\usepackage[dvipsnames]{xcolor}
\usepackage{hyperref}
\usepackage{braket}
\usepackage{bbm}
\usepackage{bm}
\usepackage{longtable}
\usepackage[bb=boondox]{mathalfa}
\usepackage{adjustbox}
\usepackage{array}
\usepackage{multirow}
\usepackage{tabularx}

\usepackage{float}

\usepackage{tikz}
\usetikzlibrary{shapes,arrows,patterns}
\usepackage{svg}

\newcommand{\dr}{\dot{\rho}}
\newcommand{\mR}{\mathcal{R}}

\newcommand{\Tr}{\operatorname{Tr}}

\newcommand{\del}{\partial}

\newcommand{\be}{\begin{equation}}
\newcommand{\ee}{\end{equation}}

\newcommand{\executeiffilenewer}[3]{%
\ifnum\pdfstrcmp{\pdffilemoddate{#1}}%
{\pdffilemoddate{#2}}>0%
{\immediate\write18{#3}}\fi%
}

\title{Spacetime as a quantum circuit}

\author[a]{A. Ramesh Chandra,}
\author[a]{Jan de Boer,}
\author[b,c]{Mario Flory,}
\author[d,1]{Michal P.\ Heller,\note{On leave of absence from: \textit{National Centre for Nuclear Research, 02-093 Warsaw, Poland}}}
\author[a]{Sergio H{\"o}rtner,}
\author[a]{and Andrew Rolph}

\affiliation[a]{Institute for Theoretical Physics, University of Amsterdam, PO Box 94485, 1090 GL Amsterdam, The Netherlands}
\affiliation[b]{Institute of Physics, Jagiellonian University, 30-348 Krak{\'o}w, Poland}
\affiliation[c]{Instituto de F{\'i}sica T{\'e}orica IFT-UAM/CSIC, Universidad Autonoma de Madrid, 28049, Madrid, Spain}
\affiliation[d]{Max Planck Institute for Gravitational Physics (Albert Einstein Institute), 14476 Potsdam-Golm, Germany}

\emailAdd{ramesh.ammanamanchi@gmail.com}
\emailAdd{j.deboer@uva.nl}
\emailAdd{mflory@th.if.uj.edu.pl}
\emailAdd{michal.p.heller@aei.mpg.de}
\emailAdd{s.hortner@uva.nl}
\emailAdd{andrew.d.rolph@googlemail.com}

\abstract{We propose that finite cutoff regions of holographic spacetimes represent quantum circuits that map between boundary states at different times and Wilsonian cutoffs, and that the complexity of those quantum circuits is given by the gravitational action. The optimal circuit minimizes the gravitational action. This is a generalization of both the ``complexity equals volume'' conjecture to unoptimized circuits, and path integral optimization to finite cutoffs. Using tools from holographic $T\bar T$, we find that surfaces of constant scalar curvature play a special role in optimizing quantum circuits. We also find an interesting connection of our proposal to kinematic space, and discuss possible circuit representations and gate counting interpretations of the gravitational action. }

\begin{document} 
\maketitle
\flushbottom

\section{Introduction} 

Quantum information theoretic concepts such as entanglement entropy 
have proven to be of fundamental importance for our
understanding of quantum gravity, most notably in the context of the AdS/CFT correspondence~\cite{Ryu:2006bv, Maldacena:1997re}.
However, it has also been claimed that ``entanglement is not enough'' \cite{Susskind:2014moa}, such as in the inability of holographic entanglement entropy to probe the late time linear growth of the Einstein-Rosen bridge in eternal black holes, and that other concepts, in particular state complexity, are needed for a more complete understanding.

The notion of state complexity in quantum mechanics refers to a setup where one is given an
initial state, a final state, a margin of error 
and a list of allowed unitary operations. The smallest number of
unitaries needed to obtain the final state from the initial state up to the margin of error is
an indication of how difficult it is to obtain the final state from the initial state. If the 
initial state is a fixed and very simple state, i.e. an unentangled product state, one can simply refer to the complexity of the final state as state complexity. 

The idea to build interesting quantum states using a limited set of
operations has been successful in condensed matter physics leading to e.g. 
tensor network representations of states~\cite{Orus:2013kga}. Moreover, 
a specific relation between tensor networks and gravity has been
proposed~\cite{Swingle:2009bg,Swingle:2012wq} in which one interprets constant time slices of AdS spacetime as a so-called multiscale entanglement
renormalization (MERA) tensor network~\cite{Vidal:2008zz}. 
As a first piece of evidence for this relation one notices that 
the tensor network in question indeed closely resembles a 
lattice discretization of an equal time slice of AdS.  If one imagines that this MERA 
tensor network is the optimal network to obtain the ground state of a (discretized) CFT, 
then the number of tensors needed equals the volume of the equal time slice. 
This subsequently led to a much more general ``complexity equals volume''
proposal~\cite{Stanford:2014jda} where one proposes that the complexity of any viable state in holographic quantum field theories
can be obtained from the minimal volume of a slice of the geometry which is anchored at 
the relevant fixed boundary time slice. Other complexity proposals include those where complexity
is computed from the action~\cite{Brown:2015bva,Brown:2015lvg} or spacetime
volumes~\cite{Couch:2016exn} evaluated in bulk Wheeler-deWitt (WdW) patches. 
All these holographic complexity proposals share
certain qualitative features that any notion of state complexity should possess, 
while still lacking a precise microscopic definition in the dual CFT. 

A continuum version of MERA~\cite{Haegeman:2011uy} was an important factor in the realization of~\cite{Jefferson:2017sdb,Chapman:2017rqy}, that a natural way to count gates and define complexity in QFT is by assigning a metric to a suitable group underlying the state preparation of interest~\cite{Nielsen1133}. While this is arguably the most promising way to define and 
prove holographic complexity proposals, or to find other gravitational manifestations of complexity, in this paper we will not directly attempt to find a precise microscopic definition of complexity in holographic CFTs. 
Instead, we will propose a significant refinement of the relation between geometry and
complexity as follows: we suggest that any spacetime region can be interpreted as a quantum
circuit, with the gravitational action providing a notion of complexity for this particular quantum circuit. 

\begin{figure}[htbp]
\centering
\includegraphics[scale=0.7]{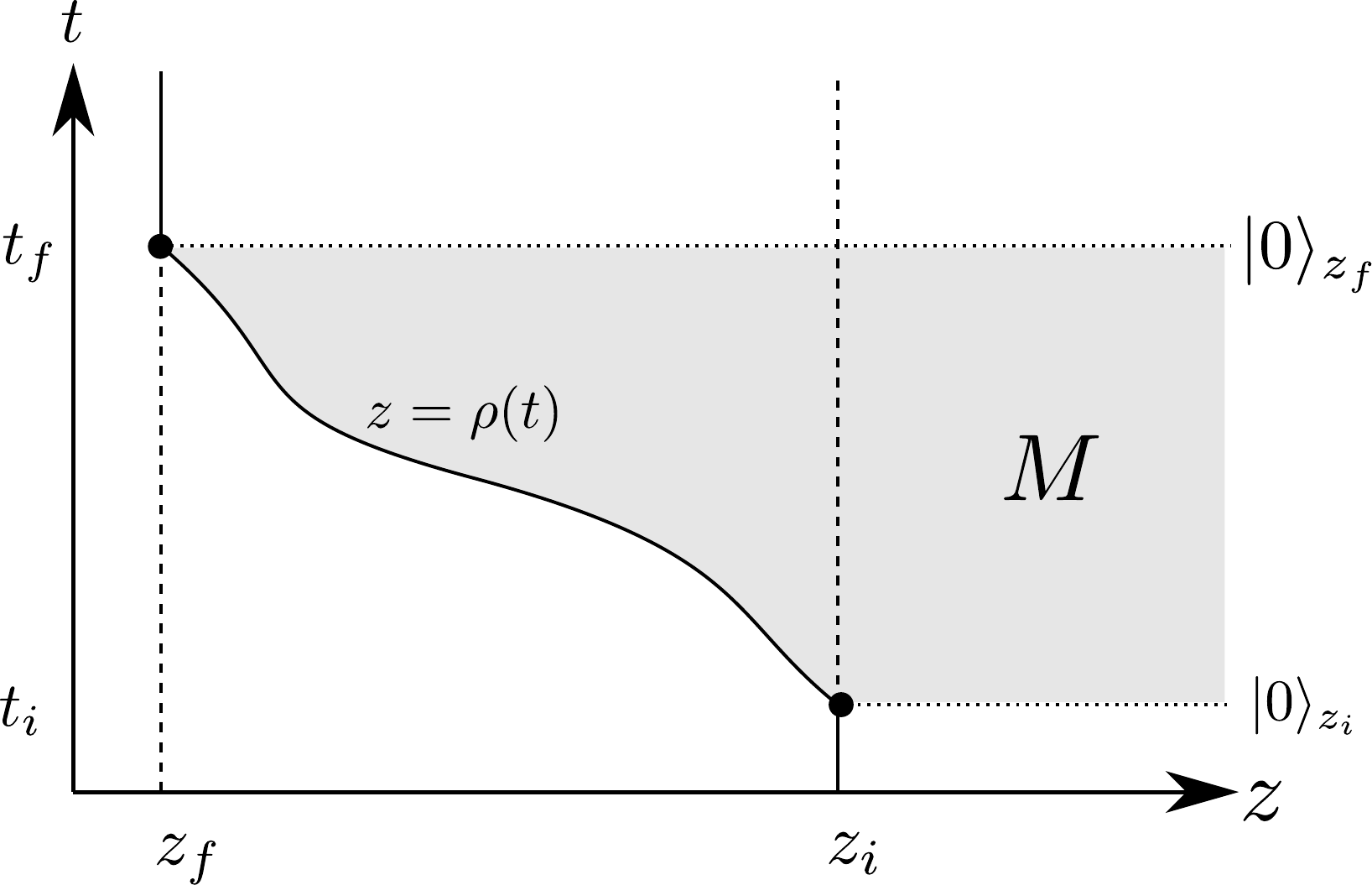}
\caption{We consider a subregion $M$ of Euclidean Poincar\'e AdS$_3$. We introduce two time-slices $t=t_{i}$ and $t=t_{f}$ corresponding to the field theory ground states $|0\rangle_{z_{i}}$ and $|0\rangle_{z_{f}}$, which are prepared for different values of the radial cutoff. The radial boundary is at finite cutoff, $z=\rho(t)$. Our proposal is that the complexity of the circuit that maps between these ground states with different finite Wilsonian cutoffs is given by the gravitational action on $M$.  }
\label{fig::bulk}
\end{figure}

Let us make our proposal more precise. Take an Euclidean asymptotically AdS geometry with radial coordinate $z$, the asymptotic
boundary being at $z=0$, and a spacetime region $M$ given by $t_i\leq t \leq t_f$
and $z\geq \rho(t)$ for some function $\rho(t)$, see figure~\ref{fig::bulk}. The AdS geometry describes the time evolution
of a given state, and the region $z\geq \rho(t)$ knows about the state, but only up to a UV
cutoff set by $\rho(t)$. Here, we use the well-known relation between the radial distance in 
AdS and the UV cutoff in the theory~\cite{Susskind:1998dq} and its recent refinement~\cite{McGough:2016lol}. The latter development links gravity in AdS spacetimes with a finite radial cutoff to finite irrelevant deformations of dual CFTs by a $T\bar{T}$ operator, and this will be a useful way of thinking about the UV cutoff in the remainder of this
paper. The bulk geometry of interest can be thought of as
describing a sequence of states, which are related to each other by both Euclidean time evolution
and a change of UV cutoff. One can ask what the complexity of this particular process is,
i.e. how many operations would be required to recover a (discretized version of) this sequence of states. For previous work combining holographic $T\bar T$ and complexity, see~\cite{Jafari:2019qns,Geng:2019yxo,Chen:2019mis} (see also~\cite{Chakraborty:2020fpt} for a related development). We propose that the number of operations, the complexity, is given by the gravitational action itself, 
evaluated with suitable boundary condition on the boundary of the spacetime region. 

Our proposal, after specializing to two boundary dimensions, and keeping only the first two orders in a Taylor expansion of the Wilsonian cutoff, coincides with the path integral optimization proposal~\cite{Caputa:2017urj,Caputa:2017yrh,Czech:2017ryf,Takayanagi:2018pml,Camargo:2019isp}, and so for holographic CFTs can be seen as generalization of that proposal to any dimension and finite cutoff. In path integral optimization one considers different preparations of a fixed state using CFT path integrals on background geometries differing by a Weyl factor. The proposal is to regard the
 change in the unnormalized path integral measure -- which for AdS${}_3$ is given by the Liouville action~\cite{Polyakov:1981rd} -- as a cost function. However, as recognized in~\cite{Camargo:2019isp}, minimization of such a cost function is not consistent with keeping only the terms which remain finite as the UV cutoff is
removed. Our proposal avoids this problem altogether by considering the full gravitational action, 
with the Liouville action merely capturing the leading two terms in the limit where
one takes the cutoff to infinity.

Coming back to our tensor network motivation, it is interesting to note that in the course 
of the past several years substantial progress has been achieved in obtaining MERA 
from a systematic coarse-graining procedure rather than using it 
merely as an efficient variational ground state ansatz for critical spin chains. 
The key idea is to employ an entanglement-based coarse-graining of the discretized Euclidean path-integral~\cite{Evenbly_2015}. This procedure is closely related to one where one puts
the corresponding conformal field theories, which arise in the continuum limit of the critical
spin chains, on curved geometries~\cite{Milsted:2018yur,Milsted:2018san}. We find the apparent connection between these ideas and our setup quite intriguing. In particular, in the language of~\cite{Milsted:2018yur} one would be tempted to call part of our circuit associated 
with moving in $t$ as composed from ``euclideons'', whereas the motion in the $z$ direction 
would have to do with ``isometries'' and possibly ``disentanglers''. 

It would be certainly very interesting to make this association more quantitative, 
perhaps using the results of~\cite{Kruthoff:2020hsi}, as there are currently 
several distinct proposals for associating geometries to MERA.
It has been suggested to connect MERA to a hyperbolic geometry of an
equal time slice of AdS as mentioned above, to a light-cone~\cite{Milsted:2018san} and to an 
auxiliary dS geometry~\cite{Czech:2015kbp} as in~\cite{Beny:2011vh}. The latter was 
motivated by work trying to probe the bulk geometry using non-local CFT observables such as 
entanglement entropy of spherical subregions, which gave rise to the kinematic space program~\cite{Czech:2015qta,Czech:2016xec,deBoer:2016pqk}. This confusing
state of affairs was one of our
motivations to try to sharpen the relation between gravity and quantum circuits. It should be also noted that a more precise relation between gravity and (other than MERA) tensor networks was proposed recently in~\cite{Bao:2018pvs,Bao:2019fpq}.

It is important to point out that we are not considering arbitrary circuits: all circuits
are essentially composed of time evolution and changes in the local cutoff starting
with a given initial state. One could certainly imagine more general circuits, but these 
would not be captured by a single semi-classical geometry and would require off-shell
gravitational configurations. The latter are typically exponentially suppressed and
we will not consider them in this paper. One can still try to find the optimal circuit
within a given semi-classical geometry, by varying over $t_i$ and over $\rho(t)$. In particular,
taking $\rho(t_i)=\infty$ corresponds to a CFT state where the CFT 
has a momentum cutoff brought down to zero, so this is akin to making the
state at $t=t_f$ from ``nothing''~\cite{Hartle:1983ai}. We find that optimizing complexity over this restricted set
of circuits gives results quite similar to other holographic complexity proposals. Perhaps
in these holographic situations, there is nothing to be gained (from a complexity point of
view) by considering circuits that involve different semiclassical geometries.

Finally, let us emphasize that there was earlier work, such as~\cite{Nozaki:2012zj,Miyaji:2015fia,Takayanagi:2018pml,Belin:2018bpg,Belin:2020oib,Boruch:2020wax,Caputa:2020fbc}, advocating for the relation between quantum circuits and holographic geometry. 
In the present work we are building on these earlier developments to bring in three important new elements into this discussion: being explicit about an initial state, realizing the need of keeping UV cut-off finite and interpreting 
it in terms of a $T\bar{T}$ deformation and, last but not least, making a connection with the kinematic space program.

The outline of this paper is as follows. In section \ref{sec.dualgravityperspective}, we will first describe the general 
setup and then give an explicit example in 
vacuum AdS${}_3$, where we will find that our notion of complexity agrees with the complexity
equals volume proposal and, in the limit $\rho(t)\rightarrow 0$,
also with path-integral optimization. Subsequently, we  discuss various finer points associated with our proposal in sections~\ref{sec::BCFT} and~\ref{sec::BCFT2}. This brings us to section \ref{sec::TTbar}, where we describe
the relation between a change in the spacetime region and $T\bar{T}$-deformations using
bulk flow equations. Considerations in this section will also allow us to argue that complexity is optimized if
the boundary of the spacetime region has constant scalar curvature. Finally, we will discuss
some ideas to more directly connect the gravitational action to a gate counting procedure in section \ref{sec::counting}, 
and end with some conclusions and suggestions for future work in section \ref{sec::outlook}.
\\

\section{Vacuum preparation using gravity \label{sec.dualgravityperspective}}

The main idea of our construction is as follows: We can produce states in a CFT using path integrals over Euclidean manifolds with a boundary and operator insertions. Similarly, path integrals over manifolds with two boundaries can be interpreted as objects mapping states to states. 
We would like to think of these path integrals as describing a circuit which prepares states or maps states to states, and associate a notion of cost function to them which measures the effort it takes to perform these CFT operations in a given way. To define such a cost function it seems inevitable to introduce some sort of cutoff in the field theory. 
This cutoff defines a local lattice spacing and provides the natural length scale at which to define tensors which make up an approximate tensor network description of the CFT operation. The cutoff could in principle be space and time dependent. 

To determine the complexity, we are going to propose to use the unnormalized CFT path integral. An important issue is how to incorporate the space and time dependent cutoff in this computation. 
In the field theory, one could try to implement this by including a space and time dependent $T\bar{T}$-deformation in the CFT which is known to implement a particular type of finite cutoff \cite{Smirnov:2016lqw,Cavaglia:2016oda}. 
It seems difficult to compute the required path integral directly in the CFT, but luckily, for CFTs with a holographic dual we can use the AdS/CFT correspondence to do the computation. 
Following~\cite{McGough:2016lol}, the relevant AdS setup is one where we move the boundary of AdS a finite distance inward, with the time (and possibly space) dependent radial position corresponding to the cutoff or coefficient of the $T\bar{T}$ deformation. 
The partition function of the CFT with cutoff is then computed, to leading order, by the on-shell value of the gravitational action with a finite instead of an asymptotic boundary.

There are some aspects of this proposal that require clarification. One is the choice of boundary for the gravitational path integral away from the surface where the cutoff CFT lives. 
For example, if the cutoff CFT lives on a hemisphere, we need to fill in the boundary of the hemisphere in AdS. There is in general no canonical slice in the bulk where the state ``lives''.
In the example that we consider below, there are always natural time-symmetric surfaces in the bulk which are the natural surfaces where to bound the bulk path integral. Another issue with the construction is whether or not to include the standard counterterms for AdS/CFT  for co-dimension one boundaries when evaluating the bulk action. Due to the existence of a finite cutoff, there is no strict need to do so, and not including them appears to be the most natural thing to do as we discuss below. 
A closely related issue comes from the fact that the full bulk region has corners, and one may need to include corner terms when evaluating the bulk action. We will also address this issue below.

\subsection{Action calculation}

For simplicity and concreteness, we are going to consider the preparation of the ground state of a 2d CFT on a line using the Euclidean path integral.
To this end, we take the standard Euclidean AdS solution, with the curvature scale $l_{AdS} = 1$,
\begin{align}
\label{eq.ads3}
ds^2 = \frac{dz^2 + dt^2 + dx^2}{z^2},   
\end{align}
and the partition function of the CFT equals the exponent of minus the on-shell bulk action
\begin{align}
\label{bulk.action}
I =  \frac{1}{\kappa}\int_{M} d^3x\, \sqrt{G} \left(R+2\right) + \frac{2}{\kappa} \int_{\del M} d^2x\, \sqrt{g}K+I_c. 
\end{align}
$M$ is the bulk region bounded by $\rho(t)\leq z \leq\infty$ and $t_i \leq t \leq t_f$, as shown in figure~\ref{fig::bulk}. The a priori finite function $\rho(z)$ interpolates between the values $z=z_{i}$ at $t=t_i$ and $z=z_{f}$ at $t=t_{f}$, with $t_i \leq t_{f}$ and $z_{f}<z_{i}$. For simplicity we also take the setup to be independent of the transverse direction $x$.  
Furthermore, we write $\kappa = 16\pi G_N$, $G$ for the 3d metric on $M$, $g$ for the induced 2d metric on $\del M$, and $K$ is the trace of the extrinsic curvature. 
$\del M$ is only piecewise smooth and has a kink or joint at $t=t_{f}$ and $t=t_i$ as shown in figure \ref{fig::bulk}. Each joint contributes a term
\begin{align}
    I_c=\frac{2}{\kappa} \int dx \sqrt{j}\ \alpha
\end{align}
to the gravitational action. Herein, $\sqrt{j}$ is the length element along the joint and $\alpha$ is simply the angle between the two normal vectors of the two surfaces coming together at the joint (which may have either sign). Joint-terms of this type were studied by Hayward in  \cite{Hayward:1993my,Brill:1994mb}, but in the Euclidean setting, which is of interest here, this was already done earlier in \cite{Hartle:1981cf}, see also the discussion in \cite{Lehner:2016vdi}.

As discussed above, we are going to interpret the on-shell value of the bulk effective action of the region $M$ as the complexity of the circuit defined by the surface $z=\rho(t)$ which maps the vacuum state $|0\rangle_{z_{i}}$ with cutoff $z_{i}$ to the vacuum state $|0\rangle_{z_{f}}$ with cutoff $z_{f}$. If we use the 
relation between a finite radial cutoff and the coefficient $\mu$ of the $T\bar{T}$ deformation 
via~\cite{McGough:2016lol},
\begin{equation}
\label{McGough}
\mu(t) = \kappa \, \rho(t)^2,
\end{equation}
we can reinterpret the states $|0\rangle_{\rho(t)}$ as ground states of the $T\bar{T}$ deformed CFT with a time-dependent coefficient $\mu(t)$.

Concretely, the induced line element on the boundary surface is 
\begin{align}
\label{eq.metricinduced}
ds^2 = \frac{(1+\dot{\rho}^2)dt^2 + dx^2}{\rho^2},   
\end{align}
its Ricci scalar reads
\begin{align}
     R^{(d-1)}=\frac{2(\rho \ddot{\rho}
     -\dr^2(1+\dr^2))}
     {(1+\dr^2)^{2}},
\end{align}
the trace of the extrinsic curvature reads
\begin{align}
     K=\frac{\rho \ddot{\rho}
     +2(1+\dr^2)}
     {(1+\dr^2)^{3/2}},
\end{align}
and from \eqref{bulk.action} we obtain 
\begin{align}
I &= \frac{-4}{\kappa}\int_{M} d^2x \int_{z=\rho}^{\infty} \frac{dz}{z^3} + \frac{2}{\kappa} \int_{\del M} d^2x\, \frac{\rho\ddot{\rho} + 2(1+\dot{\rho}^2)}{\rho^2(1+\dot{\rho}^2)}+I_c
\nonumber
\\
&= \frac{2V_x}{\kappa} \int_{t_i}^{t_{f}} dt\, \frac{  \rho\ddot{\rho}+(1+\dr^2) }{\rho^2(1+\dot{\rho}^2)}+I_c
\label{jaux22}
\end{align}
for the on-shell bulk action, where we have introduced $V_x=\int dx$. For the corner term, we also find
\begin{align}
I_c=\frac{2V_x}{\kappa}\left(\frac{\pi/2-\arctan{\dr(t_{f})}}{z_{f}}+\frac{\pi/2+\arctan{\dr(t_i)}}{z_{i}}\right).    
\end{align}
Integrating by parts, this action can be written only using first derivatives of $\rho$, yielding
\begin{align}
\label{eq.actionslim}
I = &\frac{2V_x}{\kappa} \int_{t_i}^{t_{f}} dt\, \left( \frac{1}{\rho^2}+\frac{\dot{\rho}\arctan{\dot{\rho}}}{\rho^2} \right)
+\frac{\pi V_x}{\kappa}\left(\frac{1}{z_{f}}+\frac{1}{z_{i}}\right).
\end{align}
The terms which are independent of $\rho$ do not affect the equations of motion, and can always be removed by a suitable counter term, which we will assume to be done from now on. We believe this is justified, as it is known \cite{Brill:1994mb,Lehner:2016vdi} that the joint term can spoil the additivity of the action under combining bulk regions, which besides the formulation of a well defined variational principle is usually the second main reason for adding boundary terms to the action \eqref{bulk.action}.\footnote{
Note that in our Euclidean setting, where spacelike surfaces have spacelike normal vectors, the joints under consideration are more similar to the timelike joints discussed in \cite{Brill:1994mb,Lehner:2016vdi} than spacelike ones in a Lorentzian setting. 
}

The equations of motion obtained by extremizing \eqref{eq.actionslim} read
\begin{align}
\label{eom}
    \frac{\rho \ddot{\rho}+(1+\dot{\rho}^2)}{\rho^3(1+\dot{\rho}^2)^2}=0.
\end{align}
The most immediately visible solution to this equation is the one where we formally take the limit $\dr\rightarrow\infty$. This corresponds to the boundary surface turning into an equal-time slice, which is in fact where, based on the intuition surrounding holographic complexity and tensor networks, we expect the most optimised circuit preparing the state $|0\rangle_{z_{f}}$ to live, see e.g.~\cite{Boruch:2020wax}. The generic solution to \eqref{eom} reads
\begin{align}
\label{semicirlce}
    \rho(t)=\sqrt{\mR^2-(t-t_0)^2}
\end{align}
and describes circular arcs of radius $\mR$ centered on the boundary point at $t=t_0$. The formal solution $\dr\rightarrow\infty$ corresponds to the limit of infinite radius.

Our proposal is that the Euclidean action \eqref{eq.actionslim} (excluding the $\rho$-independent remnants of the joint terms) is a measure of the complexity of preparing the state $|0\rangle_{z_{f}}$ from the state $|0\rangle_{z_{i}}$ using the circuit described by $\rho(t)$. 
The optimal circuit, with fixed Euclidean time distance $\Delta t =|t_f-t_i|$,
is then of the form (\ref{semicirlce}), and the complexity of this circuit is given by 
evaluating the Euclidean action on this solution.
With the explicit boundary conditions being $\rho(t_{f}) = z_{f}$ and $\rho(t_i)=z_{i}$, the value of the Euclidean action in the first term of \eqref{eq.actionslim} is
\begin{align}
    I = \frac{2V_x}{\kappa}\left( \frac{1}{z_{f}}\arctan{\frac{z_{i}^2-z_{f}^2 +\Delta t^2}{2z_{f}\Delta t}}
-\frac{1}{z_{i}}\arctan{\frac{z_{i}^2-z_{f}^2 -\Delta t^2}{2z_{i}\Delta t}}\right).
\end{align}
Note that this result comes entirely from the corner terms, as the first term in \eqref{jaux22} exactly vanishes on-shell. 
Interpreting it as a function of the variable $t_{i}\leq t_{f}$ while keeping $z_{i}\neq z_{f}$ fixed, we can verify that the above expression is minimized by $t_{i} = t_{f}$. This corresponds to the limit $\mR\rightarrow\infty$ or $\dr\rightarrow\infty$ and hence the equal time slice that is intuitively expected to play a special role in describing the complexity of the state $|0\rangle_{z_{f}}$. 
Using $1/\kappa = c/24$ \cite{Brown:1986nw}, the minimum value is given by 
\begin{align}
    I_{min} = \frac{c{\pi}V_x}{24}\left(\frac{1}{z_{f}}-\frac{1}{z_{i}}\right),
\end{align}
which is proportional to the spatial volume of the strip $z_{f}\leq z\leq z_{i}$ on the equal time slice at $t=t_{f}$.
Of course, if we send $z_{f}\rightarrow\epsilon\ll1$ and $z_{i}\rightarrow\infty$, this reproduces the standard result of the volume proposal  for the complexity of the CFT ground state. 
Clearly, this result also vanishes if $z_i=z_f$, which we take as a non-trivial consistency check and further justification for excluding the remnants of the joint terms in \eqref{eq.actionslim}. \footnote{As an illustrative example, imagine a Euclidean axisymmetric spacetime, with a spacetime region in the shape of a regular prism that breaks rotational symmetry around the axis to a discrete subgroup. In the limit where the radius of the prism goes to zero, the action on that region may not go to zero, as while bulk and surface terms vanish in this limit due to the vanishing of bulk volume and surface area, the joint terms will lead to a contribution proportional to an integral along the axis of symmetry. This remnant term is the analogue of the last bracket in \eqref{eq.actionslim}. }

To close this section, let us compare our results to the ones that can be obtained from the Liouville action. For $\dr\ll1$, equation \eqref{eq.actionslim} can be approximated as 
\begin{equation}
\label{eq.actionf}
I = \frac{2V_x}{\kappa} \int dt \left( \frac{1}{\rho^2} + \frac{\dot{\rho}^2}{\rho^2} \right),
\end{equation}
which, assuming no $x$-dependence, is equivalent to the Liouville Lagrangian
\begin{equation}
\label{eq.Liouville1}
S_{L} = \frac{c}{24 \pi} \int 
dt \int 
dx \left( \eta \, e^{2 \omega} + \left(\partial_{t} \omega \right)^2 + \left(\partial_{x} \omega \right)^2 \right).
\end{equation}
after a change of variables $\rho(t) \to (1/\sqrt{\eta}) \, e^{-\omega(t)}$. 
Note that the physically interesting solution $\dr\rightarrow\infty$ falls outside of the range of applicability of the approximation necessary to obtain the Liouville action from \eqref{eq.actionslim}. The equations of motion derived from \eqref{eq.actionf} take the form 
\begin{align}
\label{Liouvillerho}
\frac{\rho\ddot{\rho}+(1-\dr^2)}{\rho^3}=0.
\end{align}
As we will see below, these field equations also arise if we introduce a new time coordinate in order to bring the induced metric on the boundary into conformal gauge.

\subsection{Conformal time and extremizing the action}
\label{sec::BCFT}

There is a subtle but crucial difference between our setup discussed in the previous subsection and the calculations of 
\cite{Boruch:2020wax}, which we will discuss in this subsection in order to avoid confusion. 

In order to do so, we note that \cite{Boruch:2020wax} investigates a setup similar to the one depicted in figure \ref{fig::bulk}, and up to notation \eqref{jaux22} also appears in the appendix of that paper. Following \cite{Boruch:2020wax}, we can now introduce a conformal time $u$, with 
\begin{align}
\label{reparam}
du= \sqrt{1+\dot{\rho}(t)^2}dt,   
\end{align}
such that the line element \eqref{eq.metricinduced} is transformed into the conformal gauge form 
\begin{equation}
ds^2 = \frac{du^2 + dx^2}{\varrho(u)^2}.
\end{equation}
Here, we have introduced a new variable such that $\varrho(u(t))=\rho(t)$. Under \eqref{reparam}, the action \eqref{eq.actionslim} changes to \cite{Boruch:2020wax}
\begin{equation}
\label{eq.Iinu}
I = \frac{2 V_x}{\kappa} \int_{u_{i}[\varrho]}^{u_{f}[\varrho]}du \left( \frac{\sqrt{1-\varrho'^2}+\varrho'\arcsin{\varrho'}}{\varrho^2} \right).
\end{equation}
If we were to just identify the integrand in \eqref{eq.Iinu} as a Lagrangian and compute naively the Euler equations, we arrive at 
\begin{align}
\label{eom2}
    \frac{\varrho \varrho''+2(1-\varrho'{}^2)}{\varrho^3(1-\varrho'{}^2)^2}=0,
\end{align}
which, up to notation and the addition of a  nonzero tension term, are the equations which where studied in \cite{Boruch:2020wax}. 

The subtlety announced at the beginning of the subsection is that \eqref{reparam} is a reparametrization of time which is dependent on the variable with respect to which we want to vary the action, hence formally in going from \eqref{eq.actionslim} to \eqref{eq.Iinu} the integration bounds $u_i$ and $u_f$ become themselves functionals of $\varrho$, and will lead to a nontrivial contribution according to Leibniz's rule when varying the action. In fact it can be checked that introducing \eqref{reparam} and $\varrho(u)$ in the equation of motion \eqref{eom} gives a result 
\begin{align}
\label{Liouvillevarrho}
   \frac{\varrho\varrho''+(1-\varrho'{}^2)}{\varrho^3} =0
\end{align}
that is inequivalent to \eqref{eom2}. Interestingly, \eqref{Liouvillevarrho} has the form of the Liouville equation \eqref{Liouvillerho}, just for $\varrho(u)$ instead of $\rho(t)$.

The most commonly known example where a field-dependent reparametrization can be useful is the Lagrangian for geodesic motion, which becomes a constant when introducing affine parametrisation. Of course, this does not mean that the equations of motion degenerate, as the full information about the value of the action -- i.e.~the length of the curve -- is now entirely encoded in the integration domain. Unfortunately, the expression \eqref{eq.Iinu} rather inelegantly falls into a middle ground between the two possible extremes, as both the integrand and the integration bounds are functionals of the variable $\varrho$, and for this reason we found it intractable to work with.

This does not mean that either our work or \cite{Boruch:2020wax} are \textit{wrong}, just that we are studying a different variational problem. We work with the action \eqref{eq.actionslim} where explicitly we assume Dirichlet boundary conditions for $\rho(t)$ at the fixed values $t=t_{f}$ and $t=t_{i}$, while \cite{Boruch:2020wax} works with the action \eqref{eq.Iinu} with the implicit assumption of Dirichlet boundary conditions for the field $\varrho$ at \textit{fixed} values of $u_i,u_f$, which is an inequivalent mathematical exercise.

\subsection{Comparison to AdS/BCFT models}
\label{sec::BCFT2}

We can investigate this issue a bit further. So far, we have essentially considered what amounts to minisuperspace models, by plugging in an ansatz into the action and deriving equations of motion for the function parametrizing that ansatz, instead of first deriving general equations of motion and then simplifying them with a given ansatz. How can we write our equations of motion in a form that is more suggestive for their general meaning and potential origin? We will do this in the next
section, but as an aside, we will now demonstrate that the semicircle solutions that we found can
also be obtained if we interpret the boundary of the bulk domain as an ``end of the world 
brane'' with an energy-momentum tensor describing matter with a very specific equation of state.
The covariant equations of motion of this end of the world brane will imply the general equation
that we will derive in the next section. The derivation in the next section does not rely 
on an end of the world brane interpretation, and it remains to be seen whether this agreement
is more than a technical coincidence. 

We should also point out that
the work of \cite{Boruch:2020wax} was strongly influenced by the type of AdS/boundary CFT (BCFT) models introduced in \cite{Takayanagi:2011zk,Fujita:2011fp}. 
In such models the boundary of the space on which the BCFT lives is also extended into the bulk spacetime in the form of an end of the world brane, on which Neumann boundary conditions are imposed. 
Besides the bulk Einstein equations, this leads to an equation of motion of the form
\begin{align}
\label{Israel}
    K_{\mu\nu}-K g_{\mu\nu}=\frac{\kappa}{2}T_{\mu\nu}
\end{align}
which determines the embedding of the end of the world brane into the ambient space. These models allow for considerable bottom-up toy-model building freedom, and $T_{\mu\nu}$ is the energy-momentum tensor of any matter that lives in the brane worldvolume. In practice, it is often set to be a constant tension term
\begin{align}
\label{tension}
T_{\mu\nu}=\lambda \, g_{\mu\nu}
\end{align}
with tension $\lambda$. As reported in \cite{Boruch:2020wax}, their equation of motion is consistent with \eqref{Israel}. As we ignore tension terms, we would set the right hand side of \eqref{Israel} to zero, and apart from the equal time slice obtained by $\dr\rightarrow\infty$, our semicircular embeddings do not satisfy this equation.

Interestingly, in a Lorentzian AdS/BCFT context, semicircular embeddings into Poincar\'e AdS were derived in \cite{Erdmenger:2014xya} for a simple model of $T_{\mu\nu}$ given by a perfect fluid with equation of state $p=a\sigma$ ($p=$pressure, $\sigma$=energy density) in the limit $a\rightarrow\infty$. So we see that semicircular embeddings into a Poincar\'e AdS do satisfy an equation of the form \eqref{Israel}, just with a specific non-trivial right hand side. 
Due to the peculiar limit in the parameter $a$, $T_{\mu\nu}$ satisfies the condition 
\begin{align}
    \det[T_{\mu\nu}]=0
\end{align}
or equivalently 
\begin{align}
    T_{\mu\nu}T^{\mu\nu}-T^2=0,
\end{align}
and hence 
\begin{align}
  \det\left[  K_{\mu\nu}-K g_{\mu\nu}\right]=0,
\end{align}
respectively
\begin{align}
&(K_{\mu\nu}-K g_{\mu\nu})(K^{\mu\nu}-K g^{\mu\nu})-\Tr[K_{\mu\nu}-K g_{\mu\nu}]^2
=K_{\mu\nu}K^{\mu\nu}-K^2=0
\label{flow}
\end{align}
for our semicircular embeddings \eqref{semicirlce}, even though they were not derived from an AdS/BCFT ansatz in this paper. We will give a direct derivation of equation \eqref{flow} 
as a flow equation for our complexity proposal in the following section.

\section{Bulk action and \texorpdfstring{$T\bar T$}{TT}}
\label{sec::TTbar}

We have considered the on-shell action of a cutout region of Poincar\'e AdS$_3$, and interpreted it as a complexity functional of states in $T\bar T$-deformed holographic CFTs. The relation \eqref{McGough} between the coefficient of the $T\bar{T}$ deformation and the radial location has been derived for constant radial cutoff~\cite{McGough:2016lol,Taylor:2018xcy,Hartman:2018tkw}, but not for time-dependent $\rho(t)$. In this section we consider the flow equations which describe movement of the cutoff surface
in a fixed background. By integrating these flow equations we should be able to derive a more precise relation between the
coefficient of the $T\bar{T}$ deformation and the location of the bulk surface. In addition, these flow equations will tell
us how complexity changes as we change the surface locations, and for which surfaces complexity is optimized while keeping
the initial and final state fixed.

\subsection{Excluding counterterms} \label{sec:excluding_counterterms}

The relevant flow equation can most easily be derived using the ADM formalism \cite{ADM:1962}. We will keep the number of spacetime dimensions free in what follows, and write the metric as
\be \label{j1}
ds^2 = N^2 dr^2 + g_{\mu\nu}(x,r)(dx^{\mu} + N^{\mu} dr)(dx^{\nu} + N^{\nu} dr).
\ee
This contains the usual lapse and shift functions, for which one can locally choose a convenient gauge $N=1$ and $N^{\mu}=0$. Following ADM and choosing units so that $\kappa=1$, we now write the Lagrangian in terms of canonical variables
\be \label{j2}
{\cal L} = \sqrt{g} \left( \pi^{\mu\nu} \partial_r{g}_{\mu\nu} - N H - N^{\mu} H_{\mu}  \right),
\ee
where the lapse and shift functions appear as Lagrange multipliers enforcing the Hamiltonian and momentum constraints
\be H = H^\mu = 0. \ee 
The canonical momenta are given by \cite{Brown.York:1993}
\begin{equation} \begin{split} \label{j3}
\pi_{\mu\nu} &= \frac{1}{\sqrt{g}} \frac{\partial S}{\partial g^{\mu\nu}} \\
&= -(K_{\mu\nu} - K g_{\mu\nu})\\
&= -\frac{1}{2} \left(\partial_r g_{\mu\nu} - 
g_{\mu\nu} g^{\rho\sigma} \partial_r g_{\rho\sigma} \right),
\end{split}
\end{equation}
where in the second step we used the fact that
metric variations are given by the Brown-York tensor, and in the last step we used the explicit form of the extrinsic curvature for the metric
(\ref{j1}) in the gauge $N=1$, $N^{\mu}=0$. Of course, the same result can also be obtained by explicitly rewriting the action as in (\ref{j2}).
Using (\ref{j3}), we find for the radial derivative
\be \label{j4}
\partial_r g_{\mu\nu} = -2 \pi_{\mu\nu} + \frac{2}{d-2} g_{\mu\nu} \pi^{\rho}_{\rho}
\ee
where $d$ is the total number of bulk spacetime dimensions (in this paper we are predominantly interested in $d = 3$). The Hamiltonian constraint can be computed from (\ref{j2}) and, for unit AdS radius, one finds
\be \begin{split}
H&=R^{(d-1)} -2\Lambda - (K^2 -  K^{\mu\nu} K_{\mu\nu}) \\
&= R^{(d-1)} +(d-1)(d-2) + \pi^{\mu\nu} \pi_{\mu\nu} - \frac{1}{d-2} (\pi^{\rho}_{\rho})^2.
\end{split} \ee
It is fairly straightforward to include matter fields in the discussion; the Hamiltonian constraint will then also contain
the Hamiltonian of the matter sector, but we will for simplicity restrict to the purely gravitational case. 
To describe the flow we imagine starting with a surface at constant $r$ and moving the cutoff slightly so that $r\rightarrow r+\epsilon(x)$. For any surface, we can always locally find coordinates such that the surface is located at
fixed value of $r$ and the metric is in the ADM gauge, so there is no loss of generality in this assumption. 
Then
\be \begin{split} \label{j20b}
\delta_{\epsilon} S &= \int \epsilon(x) \partial_r g^{\mu\nu} \frac{\partial {S}}{\partial g^{\mu\nu}} \\
&= \int \sqrt{g} \epsilon(x)  \partial_r g^{\mu\nu} \pi_{\mu\nu} \\
&=  2 \int \sqrt{g}\epsilon(x) \left( {\pi}^{\mu\nu} {\pi}_{\mu\nu} - \frac{1}{d-2} ({\pi}^{\rho}_{\rho})^2\right),
\end{split} \ee
where we used equation (\ref{j4}) for the radial dependence of the metric in terms of momenta. Interestingly, this is 
precisely of $T\bar{T}$ form, but with $T$ and $\bar{T}$ defined with respect to the metric variations of the finite surface, 
not the boundary at infinity. A more coordinate independent way of stating the result is that as we move a surface in
a given AdS background, we turn on a local $T\bar{T}$-deformation with a coefficient given by the orthogonal distance between
the original and deformed surface. If we could relate the local $T$ and $\bar{T}$ on a given surface to the $T$ and $\bar{T}$ as 
defined at infinity, we could integrate these flow equations and write the final result in terms of a finite $T\bar{T}$ 
deformation of the theory at infinity. We leave a further exploration of this interesting question to future work but 
thinking of finite $T\bar{T}$ deformations in terms of a change in the boundary conditions for the metric we expect
it to involve the linearized Einstein equations around the background~\cite{Guica:2019nzm}.

Clearly, using \eqref{j3} for $d=3$ the variation of the action vanishes if equation \eqref{flow} is satisfied. 
As is clear from the Hamiltonian constraint, this condition can also be phrased as $R^{(d-1)}+(d-1)(d-2)=0$, i.e.~the boundary surface has constant scalar curvature. Therefore, to optimize the complexity of the process we should 
use constant scalar curvature surfaces; the metric on a Euclidean AdS${}_{d-1}$ manifold precisely has the required scalar curvature. 
This is consistent with the observation that complexity is minimized if we take $t_i=t_f$ and consider a purely radial surface
at the $t=t_f$ constant timeslice in section \ref{sec.dualgravityperspective}.

\subsection{Including counterterms}

So far the discussion has used the standard bulk AdS action without the inclusion of additional counterterms, which
would render the on-shell value of the action finite as one takes the surface to the asymptotic boundary. As alluded to in the 
beginning, in the original appearance of Liouville theory as defining path integral complexity, the absence of the volume counterterm
was important. Here we briefly discuss what happens if we add a volume term for the boundary surface with an
arbitrary coefficient. In our discussion of the on-shell value of the action, it would add an extra term
\be
S_{c.t} = - 2\lambda \int d^2x \sqrt{g}=
- 2\lambda \int dt dx \frac{\sqrt{1+\dot{\rho}^2}}{\rho^2}.
\ee

Adding the counterterm modifies the field equations to
\begin{align}
\label{extendedEOM}
\frac{1}{\rho^3 (1+\dot \rho^2)}\left((\rho \ddot \rho + 1+\rho^2)-\lambda \sqrt{1+\dot \rho^2}\left(\frac{1}{2}\rho \ddot \rho +1 + \dot \rho^2\right)\right)=0
\end{align}

We can also reconsider the flow equations in the presence of the volume counterterm. Denoting the volume counterterm as
\be
S_{\rm vol}=-2\lambda(d-2) \int_{\partial M} \sqrt{g}
\ee 
so that $\lambda=1$ is precisely the counterterm which would cancel the volume divergence near the AdS boundary, we now  introduce $\tilde{\pi}_{\mu\nu} = \pi_{\mu\nu} - \lambda (d-2) g_{\mu\nu}$ so that these are precisely the canonical 
momenta in the presence of the extra boundary volume term. 
The Hamiltonian constraint can be rewritten as
\be \begin{split} \label {j7}
H &= R^{(d-1)}  + (1-\lambda^2)(d-1)(d-2) + \tilde{\pi}^{\mu\nu} \tilde{\pi}_{\mu\nu} - \frac{1}{d-2} (\tilde{\pi}^{\rho}_{\rho})^2 - 2\lambda \tilde{\pi}^{\rho}_{\rho} = 0
\end{split} \ee
We can now consider two types of flows. We can consider the variation of the action as we change the radial surface in
a given background, but we can also consider the variation of the action as we perform a conformal rescaling of the metric on
the radial surface. In $d=3$, the latter does not require an adjustment of the bulk geometry, but in higher dimensions this
is no longer true. It is therefore not clear whether conformal rescalings of the induced metric on the boundary surface 
are in general compatible with keeping the initial and final states fixed in $d>3$. Regardless, the change of the action
under the first type of flow now reads
\be \begin{split} \label{j20bbb}
\delta_{\epsilon} S &= \int \epsilon(x) \partial_r g^{\mu\nu} \frac{\partial {\tilde{S}}}{\partial g^{\mu\nu}} \\
&= \int \sqrt{g} \epsilon(x)  \partial_r g^{\mu\nu} \tilde{\pi}_{\mu\nu} \\
&=  2 \int \sqrt{g}\epsilon(x) \left( {\tilde{\pi}}^{\mu\nu} {\tilde{\pi}}_{\mu\nu} - \frac{1}{d-2} ({\tilde{\pi}}^{\rho}_{\rho})^2  - \lambda  {\tilde{\pi}}^{\rho}_{\rho}   \right)
\end{split} \ee 
and for the second type of flow with $\delta g^{\mu\nu} = \epsilon(x) g^{\mu\nu}$
\be \begin{split} \label{j20}
\delta_{\epsilon} \tilde{S} &= 
\int \epsilon(x)g^{\mu\nu} \frac{\partial \tilde{S}}{\partial g^{\mu\nu}} \\
&= \int \sqrt{g} \epsilon(x) \tilde{\pi}^{\rho}_{\rho} \\
&=\frac{1}{2\lambda} \int \sqrt{g} \epsilon(x) \left( R^{(d-1)}  + (1-\lambda^2)(d-1)(d-2) + \tilde{\pi}^{\mu\nu} \tilde{\pi}_{\mu\nu} - \frac{1}{d-2} (\tilde{\pi}^{\rho}_{\rho})^2\right).
\end{split} \ee
We see that both flows take the form of $T\bar{T}$ deformations, with various extra terms such as the scalar curvature
and the trace of the stress tensor. Just as in the case without counterterm ($\lambda=0$) it would be interesting to 
integrate these flows to finite flows starting at the AdS boundary. 

The first flow is extremized when the surface obeys
\be \label{extr}
R^{(d-1)} + (d-1)(d-2) - \lambda(d-2) K =0
\ee
which still holds for an AdS${}_{d-1}$ equal time slice in AdS${}_d$. As expected, for our setup \eqref{extr} is equivalent to \eqref{extendedEOM}.
The second flow, on the other hand, is extremized when $K=\lambda(d-1)$. This does not have an extremum for an AdS${}_{d-1}$ equal time slice in AdS${}_d$ unless $\lambda=0$. 
Moreover, as we indicated above, it is not clear whether the initial state and final state are
kept fixed along the flow, and therefore the precise interpretation of this flow is somewhat unclear. 
In any case, it would be interesting to explore whether surfaces obeying (\ref{extr}) or $K=\lambda(d-1)$ 
have the potential to define a new notion of complexity.

Finally, we notice that it is also possible to add higher order counterterms, but for those the connection to $T\bar{T}$
deformations becomes more complicated.

\section{Towards counting elementary operations}
\label{sec::counting}
\subsection{Gravitational action from counting stress tensor insertions}

The bulk computation from section~\ref{sec.dualgravityperspective} and illustrated in figure~\ref{fig::bulk} can be viewed in the light of the results from the previous section as the following non-unitary circuit acting on the initial state
\be
\label{eq.circuit}
| 0 \rangle_{z_{f}} = P\exp\left[ - \int_{t_i}^{t_f} dt\, ( H_{\rho(t)} + \dot{\rho} \, [T\bar{T}]_{\rho(t)} ) \right] |0\rangle_{z_i}.
\ee
In the above expression, $H_{\rho(t)}$ represents Euclidean time evolution in a CFT with cutoff specified by $\rho(t)$, which in the bulk would correspond to moving in the $t$-direction while keeping $\rho(t)$ fixed. 
The other term represents the operator that implements a change in the scale of the theory, which we have schematically denoted by $[T\bar{T}]_{\rho(t)}$. In the bulk this would correspond to changing $\rho$ while keeping $t$ fixed. What is important is that each layer of~\eqref{eq.circuit} in general uses operators from a different theory. We assume that $H_{\rho(t)}$ and $[T\bar T ]_{\rho(t)}$ can, at least in principle, be written and understood as operators in the undeformed field theory, say by explicitly solving the Lagrangian $T\bar T$ flow equation, and that the action of those operators on states in the undeformed theory is well-defined.

Following gate counting ideas of~\cite{Nielsen1133,Jefferson:2017sdb,Chapman:2017rqy,Camargo:2019isp}, due to spatial homogeneity of our setup one might be tempted to regard $H_{\rho(t)}$ and $[T\bar{T}]_{\rho(t)}$ as two classes of elementary operations with $\rho(t)$ playing two independent roles. The first role played by $\rho(t)$ lies in labelling the elementary operations we are using (as already mentioned, both $H_{\rho(t)}$ and $[T\bar{T}]_{\rho(t)}$ are different operators for each value of $\rho(t)$). The second role stems from $dt \, \dot{\rho}$ being related to the number of times the operator $[T\bar{T}]_{\rho(t)}$ is applied in a given layer of the circuit. Correspondingly, $H_{\rho(t)}$ is applied simply $dt \, 1$ number of times.

As a result, a na\"{i}ve way of counting insertions of $H(\rho_{t})$ would be $\int_{t_{i}}^{t_{f}} dt \, 1$ and $\int_{t_{i}}^{t_{f}} dt \, | \dot{\rho}|$ when it comes to $[T\bar{T}]_{\rho(t)}$ with the total number of insertions being the sum of the two contributions. Note that since $\rho(t)$ is just a label, in principle the contribution from every layer can be weighted by some non-negative function of $\rho(t)$ -- a penalty factor that weights hardness of applications of particular transformations. The above logic was based on an $L_{1}$ norm of the vector $\{1, \dot{\rho}\}$, but in principle any norm would do.

However, our proposal is to view the action~\eqref{jaux22} or~\eqref{eq.actionslim} as a cost function for the circuit~\eqref{eq.circuit}. It seems quite straightforward to associate a suitable weight to $H_{\rho(t)}$. If we
imagine a CFT with a fixed cutoff or lattice spacing $\sim \rho$, and we count the number of lattice points in a given
Euclidean volume (which we interpret as suitable tensor operations), then we immediately obtain an answer proportional to 
$\int dt dx \rho^{-2} = \int dt V_{x} \rho^{-2}$, which is indeed proportional to the potential term in the action~\eqref{eq.actionslim}. Within the logic outlined in the previous paragraph, this corresponds to using $L_{1}$ norm with the penalty factor equal to~$V_{x} \rho^{-2}$.

The second term in the action~\eqref{eq.actionslim} is tricky to interpret within the framework of~\cite{Nielsen1133,Jefferson:2017sdb,Chapman:2017rqy}. Following the above logic, one would be naturally inclined to associate this term with the presence of $[T\bar{T}]_{\rho(t)}$ insertions in the circuit~\eqref{eq.circuit}, however, this is difficult. Writing the relevant contribution as $\frac{V_{x}\, |\arctan{\dot{\rho}}|}{\rho^{2}} \, |\dot{\rho}|$ one does not recognize a standard penalty factor in front of~$|\dot{\rho}|$ within an $L_{1}$ norm. To this end, the penalty factor is not supposed to know about what circuits do at other layers, and its dependence on $\dot{\rho}$ via $\arctan{\dot{\rho}}$ induces such a dependence. This is very much reminiscent of the discussion in~\cite{Camargo:2019isp} about viewing the Liouville action as a bona fide cost function.

Following this thread, our interpretation of the ``potential'' and ``kinetic'' terms in the gravity action~\eqref{eq.actionslim} is quite similar to an earlier discussion about the aforementioned qualitative interpretation of the Liouville action as a complexity of a tensor network~\cite{Czech:2017ryf,Caputa:2017yrh}. In our study, the ``potential'' term studies euclideons, whereas the ``kinetic'' term is associated with changes in cut-off and might be related to isometries or full layers of MERA. 

Note also that the circuit~\eqref{eq.circuit} is very similar to the expression usually written down for cMERA \cite{Haegeman:2011uy}, which takes the form of a path-ordered exponent of infinitesimal unitaries and dilatation operators. 
An immediate issue with this expression is the precise meaning of the operator $[T\bar{T}]_{\rho(t)}$ in the CFT, as we have seen that a careful construction of this operator requires one
to integrate the flow equations from the AdS boundary to the bulk surface. If we knew the precise definition
of this operator in the CFT, we could try to assign a number to this path-ordered exponent, for example by computing
the length of the trajectory in a suitable space of operators. 
It is not inconceivable that such a computation is possible, 
as we know the commutation relations between the Hamiltonian and the $T\bar{T}$ operator in the CFT, and it would be 
interesting to explore this further.

Furthermore, it is intriguing to note that $\arctan \dot{\rho}$ is the angle that the surface makes in the $z,t$-plane, so it looks like this term is measuring the amount
of effort it takes to rotate the surface in the $z,t$-plane. It would be very interesting to understand this observation better.

Since our interpretation of the gravity action in terms of gate counting is more on the qualitative side, in the following we want to propose another way of arriving at~\eqref{jaux22}.

\subsection{Relation to kinematic space}
\label{sec::kinematic}

\begin{figure}[htbp]
\centering
\includegraphics[scale=1.2]{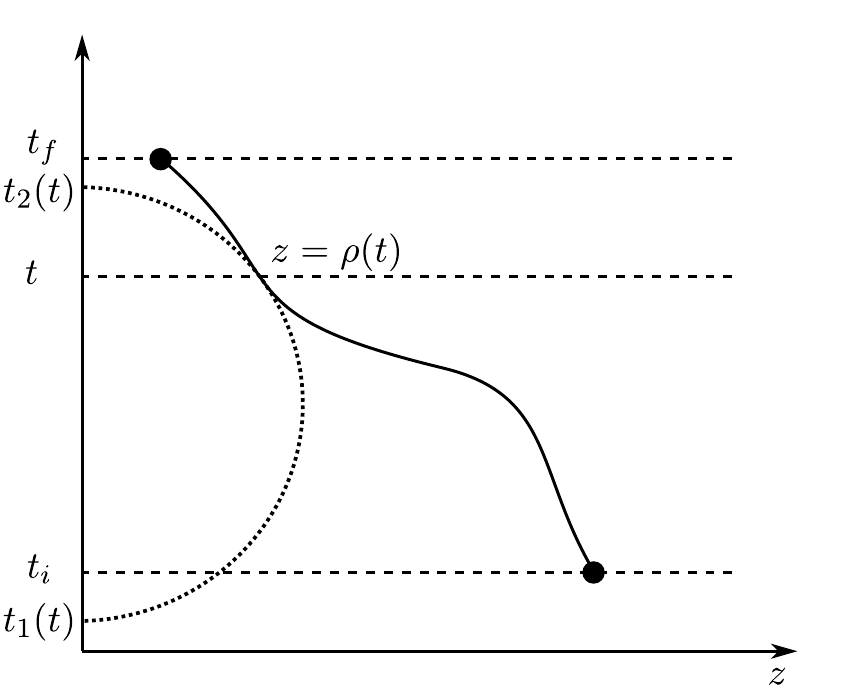}
\caption{We can parametrize a generic bulk curve $\rho(t)$ by the pairs of boundary points $(t_1(t),t_2(t))$, such that a bulk geodesic connecting these two points is tangent to the bulk curve at $z=\rho(t)$. This way, the profile $\rho(t)$ is encoded as a path in kinematic space, the space of bulk geodesics. }
\label{fig::kinematic}
\end{figure}

In the above, we have often tacitly assumed that the information about the bulk surface $z=\rho(t)$ is encoded locally in the
boundary theory. However, as our discussion of flows shows, it is highly questionable whether this is a reasonable assumption.
A better way to encode the information of the surface $z=\rho(t)$ in the boundary theory is through pairs of points
$(t_1(t),t_2(t))$ (with $x$=0) on the boundary, such that the geodesic that starts at $t_1(t)$ and ends at $t_2(t)$ is
tangent to the bulk surface at the point $(z=\rho(t),t,0)$, see figure \ref{fig::kinematic}.

This construction has the benefit of being covariant, and viewing
Euclidean time as another spatial coordinate, these geodesics encode precisely the entanglement wedges which touch the 
surface but do not cross it. In other words, they precisely encode the information about those regions of spacetime we
try to omit in our bulk path integral construction. One can ask whether there is a natural geometry associated to the
pairs of points of this type, and the answer is yes. Conformal invariance produces a natural metric on the space of pairs
of points, also known as kinematic space 
\cite{Czech:2015qta}. For the case at hand it is given up to an undetermined constant prefactor by the 2d de Sitter metric
\be \label{jaux11}
ds_{ks}^2 = \frac{-dt_1 \,dt_2}{(t_1-t_2)^2}.
\ee

In the spirit of defining complexity by assigning a metric to a group of transformations~\cite{Nielsen1133,Jefferson:2017sdb,Chapman:2017rqy}, we can now ask what the length of the path in this geometry associated with $\rho(t)$ is. 
To compute it explicitly, we need the explicit form of $t_1(t)$ and $t_2(t)$. These are given by 
\be \label{jaux12}
t_{1,2}(t) = t + \rho\dot{\rho} \pm \rho \sqrt{\dot{\rho}^2 +1}.
\ee
Consider now the action
\be
\label{actionks}
S_{ks}\sim\int \frac{dx}{\rho} ds_{ks}(t),
\ee
where we included the coordinate $x$ in units of the cutoff $\rho$, and the distance $ds$ obtained from (\ref{jaux11})
upon inserting (\ref{jaux12}). This results in 
\be
\label{kinematicspaceaction}
S_{ks} \sim \int dt dx \left| \frac{ \rho \ddot{\rho} +(1+\dot{\rho}^2 )}{\rho^2 (1+\dot{\rho}^2)} \right|,
\ee
which agrees precisely with the bulk action in the form (\ref{jaux22}) as long as $\ddot{\rho} \geq - \rho^{-1}(1+\dot{\rho}^2)$. This is related to the fact that the kinematic space is a Lorentzian manifold and the condition in question is the one that one moves there along a timelike path.

This strongly suggests that the relevant circuit geometry for these types of finite bulk surface computations is a version of kinematic space\footnote{Note that this analysis did not include corner terms contributions to the action. However, in the case of open bulk curves, in the kinematic space framework there are additional contributions that we have not included. It would be certainly interesting to see if they reproduce the corner terms. We would like to thank Bartek Czech for bringing this up.}. Note that on-shell, \eqref{kinematicspaceaction} vanishes exactly for the semi-circular arcs that solve \eqref{eom}, as they are also geodesics in AdS-space. In other words, for these solutions the path traversed in kinematic space shrinks to a point. We will come back to this point in section \ref{sec::EW}. 

Note that alternatively one may use the standard kinematic space prescription built around entanglement entropy of intervals on constant $t$ time slices. The metric~\eqref{jaux11} is the same but now with $t_1$ and $t_{2}$ replaced simply by $x_{1}$ and $x_{2}$ with
\be
x_{1,2}(t) = \pm \rho(t).
\ee
Using again~\eqref{actionks} gives this time
\be
S_{ks'}\sim \int dt dx \frac{|\dot{\rho}|}{\rho^2}.
\ee
This is clearly a different expression than~\eqref{kinematicspaceaction}, which however bears a striking similarity with the gate counting approach of~\cite{Nielsen1133,Jefferson:2017sdb,Chapman:2017rqy} when the latter uses a Manhattan norm.

Let us mention that generalizing the kinematic space consideration leading to~\eqref{kinematicspaceaction} to more complicated geometries is not obvious as minimal geodesics do not necessarily penetrate the whole spacetime. In the case of geodesics computing the entanglement entropy, these are entanglement shadows~\cite{Balasubramanian:2014sra} and they appear, for example, in the case of double-sided black holes.

Finally, let us mention that the relation between kinematic space and complexity was explored earlier in two different instances in~\cite{Abt:2017pmf} and~\cite{Chen:2020nlj}, however, these proposals are distinct from ours and use a standard entanglement-based kinematic space.

\section{Outlook}
\label{sec::outlook}

In this paper we have discussed the idea that finite spacetime regions correspond to quantum
circuits with a complexity given by the on-shell value of the gravitational action. We found 
several intriguing results, but much more work remains to be done to put our results on a
firmer footing. Perhaps the most pressing of these is to find a more precise circuit 
interpretation along the lines we discussed in the previous section. Some other obvious 
aspects to explore are the impact of counterterms, higher derivative terms and matter fields on the computations.
We list some further open issues and ideas for future work below.

\subsection*{Relation with tensor network renormalization}
As we described in the introduction, one of the long-standing questions in quantum information aspects of holography is understanding the relation between holographic geometries and MERA tensor network. With the advent of~\cite{McGough:2016lol}, it is natural to expect that the tensor network description should include $T\bar{T}$ deformations in some way. It would be certainly very interesting to explore to what extent this is the case in the existing formulation of MERA and to what extent this calls for an alternative approach, see also~\cite{Kruthoff:2020hsi}. In particular, the discussion of geometric interpretation of MERA in~\cite{Milsted:2018san} interprets latticized CFT on a hyperbolic disc geometry as intertwined layers of MERA and Euclideons (Euclidean time evolution). On the contrary, in the present paper, the hyperbolic disk geometry of the maximum volume bulk time slice arises from the absence of Euclidean time evolution in the circuit defined by~\eqref{eq.circuit}.

\subsection*{Global AdS and trivial initial state}
It is straightforward to repeat our computations in global AdS, as opposed to the Poincar\'e patch of AdS. There 
are no major conceptual changes, except that we can now choose a smooth surface without the need to pick an 
initial state. Stated differently, we have chosen a trivial initial state in the CFT with infinite cutoff, or
equivalently, we have a no-boundary type construction of the state at later times. The optimization proceeds
exactly as in Poincar\'e coordinates, and complexity is optimized if the spacetime region collapses onto an
equal time disc, with complexity proportional to the volume of the disc. 

\subsection*{Choice of time slice}
In our proposal we chose to bound our bulk spacetime subregion by constant Poincar\'e time slices. One could ask whether we could have made a different choice of boundaries and still have gotten an action that could be reasonably interepreted as circuit complexity. In the first place, our choice satisfies the physical requirement that the complexity of the circuits which do nothing (no evolution in Euclidean time, or change in cutoff) should be zero. This would not have be the case had we bounded our region by say constant $z$ surfaces rather than constant $t$. We also could have considered a wiggly spacelike boundary, which along with our timelike boundary would be dynamically determined by extremisation of the gravitational action. However, for Poincar\'e AdS we showed in section \ref{sec:excluding_counterterms} that constant $t$ slices are just such an extremum of the action, so one can consider them to have been dynamically determined from the perspective of this modified proposal. It is not clear whether this modification would always give sensible results in general asymptotically AdS spacetimes, and this is an interesting direction to consider for future work.

For stationary spacetimes, it seems reasonable to take fixed time slices, but it is
not clear what to do for more general spacetimes. States in gravity are not associated to a unique time slice. In some sense, states are associated to complete
causal diamonds in the Lorentzian signature. There is therefore no canonical choice of initial and final time slices which bound
the spacetime region.  Our proposal is to use slices with vanishing extrinsic curvature $K$, as
these are covariantly defined, and lead to a vanishing contribution of the Gibbons-Hawking boundary term. This
choice will give rise to corner contributions, but those seem unavoidable for any choice, and as we saw in the case where the
spacetime region collapses to a disc, they are a feature rather than a bug. One could, alternatively, try to extend
the spacetime region indefinitely into the past or future, and subtract the contributions of these semi-infinite
pieces later, but this procedure has exactly the same ambiguity in it. It would be interesting to have a better
understanding of the various choices one can make for the future and past boundaries and what the implications of these
choices are. It might for example also be natural to take time slices with constant scalar curvature as complexity
is locally extremized for that choice of time slice. 

\subsection*{A finite deformation of Liouville}
The effective action for a finite bounded region in AdS is of independent interest, as it computes the partition function
for the CFT with a cutoff and particular curved manifolds. In the limit where the bounded region approaches the boundary of
AdS, we recover the CFT partition function (including divergent terms), which in 2d is given by the Polyakov action, and 
in conformal gauge becomes the Liouville action. It is interesting to see that (\ref{eq.Iinu}) is apparently
a finitely deformed version of Liouville theory for a space-independent Liouville field $\varrho(u)=\exp(-\phi(u))$.
If we insert this and take $\phi(u)\rightarrow \infty$, we indeed recover Liouville theory, see also the discussion in \cite{Boruch:2020wax}. One might think that 
(\ref{eq.Iinu}) describes a finite $T\bar{T}$ deformation of Liouville theory and it would be interesting to 
make that connection precise. A possible route to address this matter is to cast the on-shell action in a form involving the scalar curvature of the cutoff surface, which seems feasible in the ADM formalism, compare with Polyakov's non-local form of the effective action, and identify the
relevant deformation.

\subsection*{Other dimensions}
In higher dimensions, the computation is more or less the same, and we will not present the relevant details here. An
exception is AdS${}_2$, where after a partial integration the action becomes proportional to $I\sim \int dt \rho^{-1}$,
which suggests that the coarse graining operation has no cost associated to it. This is perhaps a consequence of the
peculiar nature of the AdS${}_2$/CFT${}_1$ correspondence, where AdS${}_2$ is merely dual to the ground states of the
 CFT${}_1$ and is of limited relevance. It would be interesting to repeat the computation for JT gravity~\cite{Teitelboim:1983ux,Jackiw:1984je} and to compare
to flows in spaces of Hamiltonians, which are much easier to control than $T\bar{T}$ deformations in higher dimensions, 
and might lead to a more precise gate counting interpretation.

In $T\bar T$-deformed quantum mechanics, the Hamiltonian is mapped to a function of itself~\cite{Gross:2019ach,Gross:2019uxi}, 
\be \label{eq:1dTT} H \mapsto f(H). \ee
Suppose we wish to quantify the complexity of the circuit created by Euclidean time evolution,
\begin{equation} U(t) = \exp (-H\, t). \end{equation}
Given $U(t)$ of the undeformed theory, we in principle know the operator $U_f (t)$ of the deformed theory,
 \be U_f (t) = \exp (-t f(-\del_t \log U(t))), \ee 
but even given this simple relation, it is not clear how to relate the complexities of $U(t)$ and $U_f (t)$.

One puzzle arises when combining complexity and holographic $T\bar T$. Increasing the $T\bar T$ deformation is dual to bringing in the cutoff surface, which reduces the volume of the maximal boundary anchored volume slice, and by the CV conjecture would say implies that the complexity of the state similarly reduces. Assuming the volume of the maximal volume bulk slice monotonically decreases as the boundary is brought in, then this implies that the complexity is monotonically decreasing too under the flow. Is there something special about the holographic $T\bar T$ deformation such that the complexity of geometric states monotonically decreases under its flow, or is the CV proposal incorrect at finite cutoff? 

\subsection*{Lorentzian geometries}
We could repeat our computation in Lorentzian signature, but then several new features arise. First, there is the
qualitative difference of whether or not $z=\rho(t)$ describes a timelike or spacelike surface. In the timelike case
the region is delimited by the lightfronts $t=\pm z$, and the on-shell action takes the form
\begin{align}
I = \frac{2}{\kappa} \int_{\del M} d^2x\, \frac{\rho\ddot{\rho} + (1-\dot{\rho}^2)}{\rho^2(1-\dot{\rho}^2)}
\label{lorentzian.action.1}
\end{align}
Integration by parts yields
\begin{align}
I = \frac{2}{\kappa} \int_{\del M} d^2x \left[\frac{1}{\rho^2}-\frac{1}{2}\left(\text{log}(1-\dot{\rho})-\text{log}(1+\dot{\rho})\right)\right]
\label{lorentzian.action.2}
\end{align}
It is easy to see that this expression diverges in the 
limit $\dot{\rho}\rightarrow \pm 1$, i.e. when the surface is tangent to the lightfronts $t=\pm z$. It is possible to properly
define gravitational actions in the presence of null boundaries \cite{Lehner:2016vdi}, and in order for our proposal to make
sense we should modify it so that in the null limit it approaches the answer of \cite{Lehner:2016vdi}. With this modification
we would then be in agreement with the complexity equals action proposal. 

If we start with a spacelike surface and start optimizing, then there are two possibilities, we either find a constant scalar 
curvature surface, or we encounter the same null boundaries as in the previous timelike case. Which of the two optimizes the
gravitational action depends on whether we choose $+S$ of $-S$ to optimize, and since it is $e^{iS}$ which appears in the path
integral, it is not a priori clear which one of the two we should take in the absence of a precise gate counting interpretation. One would be inclined though to pick the sign such that the
term proportional to $1/\rho^2$ and independent of $\dot{\rho}$ has a positive sign so that time evolution at fixed cutoff has positive complexity. 
Regardless, we seem to universally find either constant scalar curvature surfaces or null surfaces as extrema of the extremization 
problem. 

\subsection*{BTZ black hole}
Based on general arguments, there are several key features that measures of complexity should possess, such as aforementioned asymptotic linear growth in
time in black hole backgrounds and the switchback effect~\cite{Susskind:2014jwa}. As a first heuristic check, we can investigate constant scalar curvature
slices (with the right value for the scalar curvature) in the BTZ black hole. In Kruskal coordinates, the BTZ black hole looks like
\begin{align}
ds^2 = -\frac{4\, du\,dv}{(1+uv)^2} +\frac{(1-uv)^2}{(1+uv)^2}d\phi^2
\end{align}
with the asymptotic AdS boundaries located at $uv=-1$.
The relevant constant curvature slices turn out to take the simple form $uv+ \lambda u +\mu v-1=0$. Consider the special case
$uv+(u+v)/\sinh\xi-1=0$ which intersects the boundary at $u=e^{\xi}$, $v=-e^{-\xi}$ and on the other boundary of the eternal
black hole at the point obtained by interchanging $u,v$. Shifting $\xi$ is therefore like shifting time upwards on both asymptotic
boundaries, and we are interested in the behavior at late $\xi$. The midpoint of the slice is at $u=v=\tanh\xi/2$, which indeed moves
towards the singularity at $uv=1$ as $\xi\rightarrow\infty$. Therefore, constant scalar curvature slices do correctly probe the
growing Einstein-Rosen bridge. The optimal spacetime region in this case is the region between the maximal volume slice with $K=0$ 
(which is where we propose to end the spacetime region, as discussed above) and the constant curvature slice. We have not computed the
gravitational action associated to this region but expect it to reproduce the required late time growth. As the maximal volume slice is
also explicitly computable~\cite{Carmi:2017jqz}, we leave this interesting exercise to future work.

\subsection*{Higher curvature corrections}

We have proposed that the complexity of the circuit that maps between ground states in two EFTs with different finite cutoffs is given by the on-shell gravitational action. Considering the effect of higher curvature corrections on the gravitational action, and therefore complexity, would be a natural extension to this proposal. Higher curvature corrections to the holographic complexity=volume proposal were recently studied in~\cite{Hernandez:2020nem}. The simplest example to study is Gauss-Bonnet (GB) gravity in AdS$_5$. The GB correction
\begin{equation}
\mathcal{L}^{GB} = R^2 - 4 R_{\mu\nu}R^{\mu\nu} + R_{\mu\nu\rho\sigma}R^{\mu\nu\rho\sigma} 
\end{equation}
is a constant in the vacuum AdS geometry we have studied in this work, so simply rescales the contribution from the volume of $M$ to the on-shell action. In general, however, we expect non-trivial contributions from the correction to the boundary Lagrangian, and from the bulk Lagrangian $\mathcal{L}^{GB}$ for perturbed geometries. Seeing whether the resulting on-shell action could be interpreted as the complexity of a circuit in $T\bar T$-deformed CFT$_4$ with $a \neq c$ would be interesting. 

To our knowledge the generalization of holographic $T\bar T$ to higher curvature gravity has not been studied. There is general method to determine the deformation to a holographic CFT needed to put the dual gravity theory at finite cutoff~\cite{Taylor:2018xcy,Hartman:2018tkw}. To start we note that in a theory with only one dimensionful parameter $\mu$ (assuming it to have no spacetime dependence) such as a $T\bar T$ deformed CFT, the effective action changes under infinitesimal length rescaling as
\begin{equation} \label{eq:flow} \Delta_\mu \mu \del_\mu W = \int d^d x \sqrt{\gamma} \langle \Tr T \rangle \end{equation}
where $\Delta_\mu$ is the scaling dimension of $\mu$. The flow of the boundary action under changing $\mu$ is thus determined by the trace of the boundary stress tensor, which is related to the bulk Brown-York stress tensor through $T_{ij} = r_c^{d-2} \tilde T_{ij}$. The essence of the method is to use the Hamiltonian constraint to  eliminate extrinsic curvature terms in the Brown-York stress tensor appearing in \eqref{eq:flow}. Following this procedure for pure Einstein gravity gives the flow equation
\begin{equation} \label{eq:def}
\frac{\del W}{\del \mu} = \int d^d x \sqrt{\gamma} \left(T^{ij}T_{ij} - \frac{1}{d-1}(T^i_i )^2 \right )
\end{equation}
This is the field theory deformation flow equation for Einstein gravity at finite cutoff. When we add higher curvature corrections, the method of substituting out extrinsic curvature terms in the Brown-York stress tensor for powers of the stress tensor using the Hamiltonian constraint does not change, but the Brown-York stress tensor and the radial Hamiltonian constraint do~\cite{Davis:2002gn}, and the final result will differ from \eqref{eq:def}.

\subsection*{Relation to entanglement wedge reconstruction}
\label{sec::EW}

Finally, it is interesting to ask whether it is a mathematical coincidence or not that the solutions \eqref{semicirlce} to \eqref{eom} are semicircular arcs, just like geodesics in the Poincar\'e AdS-background, even though in our setup they describe the embeddings of co-dimension one surfaces, not entanglement entropy. 
In section \ref{sec::kinematic}, we addressed this question from a kinematic space perspective.  There, we showed that a generic bulk profile $\rho(t)$ can as well be described as a path in kinematic space (neglecting a possible $x$-dependence as throughout the paper), but equations \eqref{eom} and \eqref{semicirlce} enforce this path to shrink down to a length zero point for fixed boundary conditions $t_i, t_f, z_i, z_f$. 

Assuming that our results can be generalised to the Lorentzian case, it will be interesting to investigate whether these semicircular embeddings are a consequence of entanglement wedge reconstruction or entanglement wedge nesting \cite{Wall:2012uf,Czech:2012bh,Akers:2016ugt}. For example, one might imagine that the optimization of the path integral (subject to the boundary conditions $t_i, t_f, z_i, z_f$) drives the bulk surface as deep into the bulk as possible until it leaves the entanglement wedge and cannot move further. In the fully Lorentzian case, it would be interesting to determine whether this only reproduces the extremal surface at the edge of the entanglement wedge, or also its null boundaries. At least extremal area surfaces are expected to play a role of special importance in quantum gravity for quite generic reasons \cite{Camps:2019opl}.    As said in section \ref{sec::BCFT}, such semicircular boundaries have also been found in \cite{Erdmenger:2014xya} as solutions of an AdS/BCFT toy-model with non-trivial matter content living in the worldvolume of the end of the world brane. In that paper, it was shown as a consequence of physical energy conditions on the worldvolume matter fields that the corresponding branes generally have to be  \textit{extremal surface barriers} in the sense of \cite{Engelhardt:2013tra}, and the branes with semicircular embedding profile where precisely the ones staying as close to the boundary as possible without violating this condition.   

While all these different observations seem to point to a nontrivial quantum information theoretic reason for the embeddings \eqref{semicirlce} being the correct ones, we leave further investigation of this to future work. Besides extending our work to the Lorentzian case, investigating similar setups in higher dimensions or on nontrivial backgrounds such as BTZ may yield further insight.

\acknowledgments
We would like to thank Shira Chapman and Ignacio Reyes for being involved in the initial part of this collaboration and Bartek Czech for useful discussions. The Gravity, Quantum Fields and Information (GQFI) group at AEI is supported by the Alexander von Humboldt Foundation and the Federal Ministry for Education and Research through the Sofja Kovalevskaja Award. AR is supported by the Stichting Nederlandse Wetenschappelijk Onderzoek Instituten (NWO-I). 
JdB is supported by the European Research Council
under the European Unions Seventh Framework Programme (FP7/2007-2013), ERC Grant agreement ADG
834878. 
The work of MF was supported by the Polish National Science Centre (NCN) grant 2017/24/C/ST2/00469 until November 16th 2020 and through the grants SEV-2016-0597 and PGC2018-095976-B-C21 from MCIU/AEI/FEDER, UE since December 1st 2020. RC acknowledges support from the Netherlands Organisation for Scientific Research (NWO-I). SH holds a fellowship from the Ramon Areces Foundation (Spain).

\bibliography{references}

\providecommand{\href}[2]{#2}\begingroup\raggedright\begin{thebibliography}{10}

\bibitem{Ryu:2006bv}
S.~Ryu and T.~Takayanagi, {\it {Holographic derivation of entanglement entropy
  from AdS/CFT}},  {\em Phys. Rev. Lett.} {\bf 96} (2006) 181602,
  [\href{http://arxiv.org/abs/hep-th/0603001}{{\tt hep-th/0603001}}].

\bibitem{Maldacena:1997re}
J.~M. Maldacena, {\it {The Large N limit of superconformal field theories and
  supergravity}},  {\em Int.J.Theor.Phys. 38 1113-1133 and also
  Adv.Theor.Math.Phys. 2 231-252} (1999 and 1998)
  [\href{http://arxiv.org/abs/hep-th/9711200}{{\tt hep-th/9711200}}].

\bibitem{Susskind:2014moa}
L.~Susskind, {\it {Entanglement is not enough}},  {\em Fortsch. Phys.} {\bf 64}
  (2016) 49--71, [\href{http://arxiv.org/abs/1411.0690}{{\tt
  arXiv:1411.0690}}].

\bibitem{Orus:2013kga}
R.~Orus, {\it {A Practical Introduction to Tensor Networks: Matrix Product
  States and Projected Entangled Pair States}},  {\em Annals Phys.} {\bf 349}
  (2014) 117--158, [\href{http://arxiv.org/abs/1306.2164}{{\tt
  arXiv:1306.2164}}].

\bibitem{Swingle:2009bg}
B.~Swingle, {\it {Entanglement Renormalization and Holography}},  {\em Phys.
  Rev. D} {\bf 86} (2012) 065007, [\href{http://arxiv.org/abs/0905.1317}{{\tt
  arXiv:0905.1317}}].

\bibitem{Swingle:2012wq}
B.~Swingle, {\it {Constructing holographic spacetimes using entanglement
  renormalization}},  \href{http://arxiv.org/abs/1209.3304}{{\tt
  arXiv:1209.3304}}.

\bibitem{Vidal:2008zz}
G.~Vidal, {\it {Class of Quantum Many-Body States That Can Be Efficiently
  Simulated}},  {\em Phys. Rev. Lett.} {\bf 101} (2008) 110501,
  [\href{http://arxiv.org/abs/quant-ph/0610099}{{\tt quant-ph/0610099}}].

\bibitem{Stanford:2014jda}
D.~Stanford and L.~Susskind, {\it {Complexity and Shock Wave Geometries}},
  {\em Phys. Rev. D} {\bf 90} (2014), no.~12 126007,
  [\href{http://arxiv.org/abs/1406.2678}{{\tt arXiv:1406.2678}}].

\bibitem{Brown:2015bva}
A.~R. Brown, D.~A. Roberts, L.~Susskind, B.~Swingle, and Y.~Zhao, {\it
  {Holographic Complexity Equals Bulk Action?}},  {\em Phys. Rev. Lett.} {\bf
  116} (2016), no.~19 191301, [\href{http://arxiv.org/abs/1509.07876}{{\tt
  arXiv:1509.07876}}].

\bibitem{Brown:2015lvg}
A.~R. Brown, D.~A. Roberts, L.~Susskind, B.~Swingle, and Y.~Zhao, {\it
  {Complexity, action, and black holes}},  {\em Phys. Rev. D} {\bf 93} (2016),
  no.~8 086006, [\href{http://arxiv.org/abs/1512.04993}{{\tt
  arXiv:1512.04993}}].

\bibitem{Couch:2016exn}
J.~Couch, W.~Fischler, and P.~H. Nguyen, {\it {Noether charge, black hole
  volume, and complexity}},  {\em JHEP} {\bf 03} (2017) 119,
  [\href{http://arxiv.org/abs/1610.02038}{{\tt arXiv:1610.02038}}].

\bibitem{Haegeman:2011uy}
J.~Haegeman, T.~J. Osborne, H.~Verschelde, and F.~Verstraete, {\it
  {Entanglement Renormalization for Quantum Fields in Real Space}},  {\em Phys.
  Rev. Lett.} {\bf 110} (2013), no.~10 100402,
  [\href{http://arxiv.org/abs/1102.5524}{{\tt arXiv:1102.5524}}].

\bibitem{Jefferson:2017sdb}
R.~Jefferson and R.~C. Myers, {\it {Circuit complexity in quantum field
  theory}},  {\em JHEP} {\bf 10} (2017) 107,
  [\href{http://arxiv.org/abs/1707.08570}{{\tt arXiv:1707.08570}}].

\bibitem{Chapman:2017rqy}
S.~Chapman, M.~P. Heller, H.~Marrochio, and F.~Pastawski, {\it {Toward a
  Definition of Complexity for Quantum Field Theory States}},  {\em Phys. Rev.
  Lett.} {\bf 120} (2018), no.~12 121602,
  [\href{http://arxiv.org/abs/1707.08582}{{\tt arXiv:1707.08582}}].

\bibitem{Nielsen1133}
M.~A. Nielsen, M.~R. Dowling, M.~Gu, and A.~C. Doherty, {\it Quantum
  computation as geometry},  {\em Science} {\bf 311} (2006), no.~5764
  1133--1135, [\href{http://arxiv.org/abs/quant-ph/0603161}{{\tt
  quant-ph/0603161}}].

\bibitem{Susskind:1998dq}
L.~Susskind and E.~Witten, {\it {The Holographic bound in anti-de Sitter
  space}},  \href{http://arxiv.org/abs/hep-th/9805114}{{\tt hep-th/9805114}}.

\bibitem{McGough:2016lol}
L.~McGough, M.~Mezei, and H.~Verlinde, {\it {Moving the CFT into the bulk with
  $ T\overline{T} $}},  {\em JHEP} {\bf 04} (2018) 010,
  [\href{http://arxiv.org/abs/1611.03470}{{\tt arXiv:1611.03470}}].

\bibitem{Jafari:2019qns}
G.~Jafari, A.~Naseh, and H.~Zolfi, {\it {Path Integral Optimization for
  $T\bar{T}$ Deformation}},  {\em Phys. Rev. D} {\bf 101} (2020), no.~2 026007,
  [\href{http://arxiv.org/abs/1909.02357}{{\tt arXiv:1909.02357}}].

\bibitem{Geng:2019yxo}
H.~Geng, {\it {$T\bar{T}$ Deformation and the Complexity=Volume Conjecture}},
  {\em Fortsch. Phys.} {\bf 68} (2020), no.~7 2000036,
  [\href{http://arxiv.org/abs/1910.08082}{{\tt arXiv:1910.08082}}].

\bibitem{Chen:2019mis}
B.~Chen, L.~Chen, and C.-Y. Zhang, {\it {Surface/state correspondence and
  $T\overline{T}$ deformation}},  {\em Phys. Rev. D} {\bf 101} (2020), no.~10
  106011, [\href{http://arxiv.org/abs/1907.12110}{{\tt arXiv:1907.12110}}].

\bibitem{Chakraborty:2020fpt}
S.~Chakraborty, G.~Katoch, and S.~R. Roy, {\it {Holographic Complexity of LST
  and Single Trace $T\bar{T}$}},  \href{http://arxiv.org/abs/2012.11644}{{\tt
  arXiv:2012.11644}}.

\bibitem{Caputa:2017urj}
P.~Caputa, N.~Kundu, M.~Miyaji, T.~Takayanagi, and K.~Watanabe, {\it {Anti-de
  Sitter Space from Optimization of Path Integrals in Conformal Field
  Theories}},  {\em Phys. Rev. Lett.} {\bf 119} (2017), no.~7 071602,
  [\href{http://arxiv.org/abs/1703.00456}{{\tt arXiv:1703.00456}}].

\bibitem{Caputa:2017yrh}
P.~Caputa, N.~Kundu, M.~Miyaji, T.~Takayanagi, and K.~Watanabe, {\it {Liouville
  Action as Path-Integral Complexity: From Continuous Tensor Networks to
  AdS/CFT}},  {\em JHEP} {\bf 11} (2017) 097,
  [\href{http://arxiv.org/abs/1706.07056}{{\tt arXiv:1706.07056}}].

\bibitem{Czech:2017ryf}
B.~Czech, {\it {Einstein Equations from Varying Complexity}},  {\em Phys. Rev.
  Lett.} {\bf 120} (2018), no.~3 031601,
  [\href{http://arxiv.org/abs/1706.00965}{{\tt arXiv:1706.00965}}].

\bibitem{Takayanagi:2018pml}
T.~Takayanagi, {\it {Holographic Spacetimes as Quantum Circuits of
  Path-Integrations}},  {\em JHEP} {\bf 12} (2018) 048,
  [\href{http://arxiv.org/abs/1808.09072}{{\tt arXiv:1808.09072}}].

\bibitem{Camargo:2019isp}
H.~A. Camargo, M.~P. Heller, R.~Jefferson, and J.~Knaute, {\it {Path integral
  optimization as circuit complexity}},  {\em Phys. Rev. Lett.} {\bf 123}
  (2019), no.~1 011601, [\href{http://arxiv.org/abs/1904.02713}{{\tt
  arXiv:1904.02713}}].

\bibitem{Polyakov:1981rd}
A.~M. Polyakov, {\it {Quantum Geometry of Bosonic Strings}},  {\em Phys. Lett.
  B} {\bf 103} (1981) 207--210.

\bibitem{Evenbly_2015}
G.~Evenbly and G.~Vidal, {\it Tensor network renormalization yields the
  multiscale entanglement renormalization ansatz},  {\em Physical Review
  Letters} {\bf 115} (Nov, 2015) [\href{http://arxiv.org/abs/1502.05385}{{\tt
  arXiv:1502.05385}}].

\bibitem{Milsted:2018yur}
A.~Milsted and G.~Vidal, {\it {Tensor networks as path integral geometry}},
  \href{http://arxiv.org/abs/1807.02501}{{\tt arXiv:1807.02501}}.

\bibitem{Milsted:2018san}
A.~Milsted and G.~Vidal, {\it {Geometric interpretation of the multi-scale
  entanglement renormalization ansatz}},
  \href{http://arxiv.org/abs/1812.00529}{{\tt arXiv:1812.00529}}.

\bibitem{Kruthoff:2020hsi}
J.~Kruthoff and O.~Parrikar, {\it {On the flow of states under
  $T\overline{T}$}},  \href{http://arxiv.org/abs/2006.03054}{{\tt
  arXiv:2006.03054}}.

\bibitem{Czech:2015kbp}
B.~Czech, L.~Lamprou, S.~McCandlish, and J.~Sully, {\it {Tensor Networks from
  Kinematic Space}},  {\em JHEP} {\bf 07} (2016) 100,
  [\href{http://arxiv.org/abs/1512.01548}{{\tt arXiv:1512.01548}}].

\bibitem{Beny:2011vh}
C.~Beny, {\it {Causal structure of the entanglement renormalization ansatz}},
  {\em New J. Phys.} {\bf 15} (2013) 023020,
  [\href{http://arxiv.org/abs/1110.4872}{{\tt arXiv:1110.4872}}].

\bibitem{Czech:2015qta}
B.~Czech, L.~Lamprou, S.~McCandlish, and J.~Sully, {\it {Integral Geometry and
  Holography}},  {\em JHEP} {\bf 10} (2015) 175,
  [\href{http://arxiv.org/abs/1505.05515}{{\tt arXiv:1505.05515}}].

\bibitem{Czech:2016xec}
B.~Czech, L.~Lamprou, S.~McCandlish, B.~Mosk, and J.~Sully, {\it {A
  Stereoscopic Look into the Bulk}},  {\em JHEP} {\bf 07} (2016) 129,
  [\href{http://arxiv.org/abs/1604.03110}{{\tt arXiv:1604.03110}}].

\bibitem{deBoer:2016pqk}
J.~de~Boer, F.~M. Haehl, M.~P. Heller, and R.~C. Myers, {\it {Entanglement,
  holography and causal diamonds}},  {\em JHEP} {\bf 08} (2016) 162,
  [\href{http://arxiv.org/abs/1606.03307}{{\tt arXiv:1606.03307}}].

\bibitem{Bao:2018pvs}
N.~Bao, G.~Penington, J.~Sorce, and A.~C. Wall, {\it {Beyond Toy Models:
  Distilling Tensor Networks in Full AdS/CFT}},  {\em JHEP} {\bf 11} (2019)
  069, [\href{http://arxiv.org/abs/1812.01171}{{\tt arXiv:1812.01171}}].

\bibitem{Bao:2019fpq}
N.~Bao, G.~Penington, J.~Sorce, and A.~C. Wall, {\it {Holographic Tensor
  Networks in Full AdS/CFT}},  \href{http://arxiv.org/abs/1902.10157}{{\tt
  arXiv:1902.10157}}.

\bibitem{Hartle:1983ai}
J.~Hartle and S.~Hawking, {\it {Wave Function of the Universe}},  {\em Adv.
  Ser. Astrophys. Cosmol.} {\bf 3} (1987) 174--189.

\bibitem{Nozaki:2012zj}
M.~Nozaki, S.~Ryu, and T.~Takayanagi, {\it {Holographic Geometry of
  Entanglement Renormalization in Quantum Field Theories}},  {\em JHEP} {\bf
  10} (2012) 193, [\href{http://arxiv.org/abs/1208.3469}{{\tt
  arXiv:1208.3469}}].

\bibitem{Miyaji:2015fia}
M.~Miyaji, T.~Numasawa, N.~Shiba, T.~Takayanagi, and K.~Watanabe, {\it
  {Continuous Multiscale Entanglement Renormalization Ansatz as Holographic
  Surface-State Correspondence}},  {\em Phys. Rev. Lett.} {\bf 115} (2015),
  no.~17 171602, [\href{http://arxiv.org/abs/1506.01353}{{\tt
  arXiv:1506.01353}}].

\bibitem{Belin:2018bpg}
A.~Belin, A.~Lewkowycz, and G.~S\'arosi, {\it {Complexity and the bulk volume,
  a new York time story}},  {\em JHEP} {\bf 03} (2019) 044,
  [\href{http://arxiv.org/abs/1811.03097}{{\tt arXiv:1811.03097}}].

\bibitem{Belin:2020oib}
A.~Belin, A.~Lewkowycz, and G.~Sarosi, {\it {Gravitational path integral from
  the $T^2$ deformation}},  {\em JHEP} {\bf 09} (2020) 156,
  [\href{http://arxiv.org/abs/2006.01835}{{\tt arXiv:2006.01835}}].

\bibitem{Boruch:2020wax}
J.~Boruch, P.~Caputa, and T.~Takayanagi, {\it {Path-Integral Optimization from
  Hartle-Hawking Wave Function}},  \href{http://arxiv.org/abs/2011.08188}{{\tt
  arXiv:2011.08188}}.

\bibitem{Caputa:2020fbc}
P.~Caputa, J.~Kruthoff, and O.~Parrikar, {\it {Building Tensor Networks for
  Holographic States}},  \href{http://arxiv.org/abs/2012.05247}{{\tt
  arXiv:2012.05247}}.

\bibitem{Smirnov:2016lqw}
F.~Smirnov and A.~Zamolodchikov, {\it {On space of integrable quantum field
  theories}},  {\em Nucl. Phys. B} {\bf 915} (2017) 363--383,
  [\href{http://arxiv.org/abs/1608.05499}{{\tt arXiv:1608.05499}}].

\bibitem{Cavaglia:2016oda}
A.~Cavagli\`a, S.~Negro, I.~M. Sz\'ecs\'enyi, and R.~Tateo, {\it {$T
  \bar{T}$-deformed 2D Quantum Field Theories}},  {\em JHEP} {\bf 10} (2016)
  112, [\href{http://arxiv.org/abs/1608.05534}{{\tt arXiv:1608.05534}}].

\bibitem{Hayward:1993my}
G.~Hayward, {\it {Gravitational action for space-times with nonsmooth
  boundaries}},  {\em Phys. Rev. D} {\bf 47} (1993) 3275--3280.

\bibitem{Brill:1994mb}
D.~Brill and G.~Hayward, {\it {Is the gravitational action additive?}},  {\em
  Phys. Rev. D} {\bf 50} (1994) 4914--4919,
  [\href{http://arxiv.org/abs/gr-qc/9403018}{{\tt gr-qc/9403018}}].

\bibitem{Hartle:1981cf}
J.~Hartle and R.~Sorkin, {\it {Boundary Terms in the Action for the Regge
  Calculus}},  {\em Gen. Rel. Grav.} {\bf 13} (1981) 541--549.

\bibitem{Lehner:2016vdi}
L.~Lehner, R.~C. Myers, E.~Poisson, and R.~D. Sorkin, {\it {Gravitational
  action with null boundaries}},  {\em Phys. Rev. D} {\bf 94} (2016), no.~8
  084046, [\href{http://arxiv.org/abs/1609.00207}{{\tt arXiv:1609.00207}}].

\bibitem{Brown:1986nw}
J.~Brown and M.~Henneaux, {\it {Central Charges in the Canonical Realization of
  Asymptotic Symmetries: An Example from Three-Dimensional Gravity}},  {\em
  Commun. Math. Phys.} {\bf 104} (1986) 207--226.

\bibitem{Takayanagi:2011zk}
T.~Takayanagi, {\it {Holographic Dual of BCFT}},  {\em Phys. Rev. Lett.} {\bf
  107} (2011) 101602, [\href{http://arxiv.org/abs/1105.5165}{{\tt
  arXiv:1105.5165}}].

\bibitem{Fujita:2011fp}
M.~Fujita, T.~Takayanagi, and E.~Tonni, {\it {Aspects of AdS/BCFT}},  {\em
  JHEP} {\bf 11} (2011) 043, [\href{http://arxiv.org/abs/1108.5152}{{\tt
  arXiv:1108.5152}}].

\bibitem{Erdmenger:2014xya}
J.~Erdmenger, M.~Flory, and M.-N. Newrzella, {\it {Bending branes for DCFT in
  two dimensions}},  {\em JHEP} {\bf 01} (2015) 058,
  [\href{http://arxiv.org/abs/1410.7811}{{\tt arXiv:1410.7811}}].

\bibitem{Taylor:2018xcy}
M.~Taylor, {\it {TT deformations in general dimensions}},
  \href{http://arxiv.org/abs/1805.10287}{{\tt arXiv:1805.10287}}.

\bibitem{Hartman:2018tkw}
T.~Hartman, J.~Kruthoff, E.~Shaghoulian, and A.~Tajdini, {\it {Holography at
  finite cutoff with a $T^2$ deformation}},  {\em JHEP} {\bf 03} (2019) 004,
  [\href{http://arxiv.org/abs/1807.11401}{{\tt arXiv:1807.11401}}].

\bibitem{ADM:1962}
R.~Arnowitt, S.~Deser, and W.~Misner, {\it {The dynamics of General
  Relativity}},  in {\em Gravitation: an introduction to current research}
  (L.~Witten, ed.), ch.~7, pp.~227--265.
\newblock Wiley, New York, 1962.

\bibitem{Brown.York:1993}
J.~D. Brown and J.~W. York, {\it {Quasilocal energy and conserved charges
  derived from the gravitational action}},  {\em Phys. Rev. D} {\bf 47} (1993)
  1407--1419.

\bibitem{Guica:2019nzm}
M.~Guica and R.~Monten, {\it {$T\bar T$ and the mirage of a bulk cutoff}},
  \href{http://arxiv.org/abs/1906.11251}{{\tt arXiv:1906.11251}}.

\bibitem{Balasubramanian:2014sra}
V.~Balasubramanian, B.~D. Chowdhury, B.~Czech, and J.~de~Boer, {\it
  {Entwinement and the emergence of spacetime}},  {\em JHEP} {\bf 01} (2015)
  048, [\href{http://arxiv.org/abs/1406.5859}{{\tt arXiv:1406.5859}}].

\bibitem{Abt:2017pmf}
R.~Abt, J.~Erdmenger, H.~Hinrichsen, C.~M. Melby-Thompson, R.~Meyer, C.~Northe,
  and I.~A. Reyes, {\it {Topological Complexity in AdS$_3$/CFT$_2$}},  {\em
  Fortsch. Phys.} {\bf 66} (2018), no.~6 1800034,
  [\href{http://arxiv.org/abs/1710.01327}{{\tt arXiv:1710.01327}}].

\bibitem{Chen:2020nlj}
B.~Chen, B.~Czech, and Z.-z. Wang, {\it {Cutoff Dependence and Complexity of
  the CFT$_2$ Ground State}},  \href{http://arxiv.org/abs/2004.11377}{{\tt
  arXiv:2004.11377}}.

\bibitem{Teitelboim:1983ux}
C.~Teitelboim, {\it {Gravitation and Hamiltonian Structure in Two Space-Time
  Dimensions}},  {\em Phys. Lett. B} {\bf 126} (1983) 41--45.

\bibitem{Jackiw:1984je}
R.~Jackiw, {\it {Lower Dimensional Gravity}},  {\em Nucl. Phys. B} {\bf 252}
  (1985) 343--356.

\bibitem{Gross:2019ach}
D.~J. Gross, J.~Kruthoff, A.~Rolph, and E.~Shaghoulian, {\it {$T\overline{T}$
  in AdS$_2$ and Quantum Mechanics}},  {\em Phys. Rev. D} {\bf 101} (2020),
  no.~2 026011, [\href{http://arxiv.org/abs/1907.04873}{{\tt
  arXiv:1907.04873}}].

\bibitem{Gross:2019uxi}
D.~J. Gross, J.~Kruthoff, A.~Rolph, and E.~Shaghoulian, {\it {Hamiltonian
  deformations in quantum mechanics, $T\bar T$, and the SYK model}},  {\em
  Phys. Rev. D} {\bf 102} (2020), no.~4 046019,
  [\href{http://arxiv.org/abs/1912.06132}{{\tt arXiv:1912.06132}}].

\bibitem{Susskind:2014jwa}
L.~Susskind and Y.~Zhao, {\it {Switchbacks and the Bridge to Nowhere}},
  \href{http://arxiv.org/abs/1408.2823}{{\tt arXiv:1408.2823}}.

\bibitem{Carmi:2017jqz}
D.~Carmi, S.~Chapman, H.~Marrochio, R.~C. Myers, and S.~Sugishita, {\it {On the
  Time Dependence of Holographic Complexity}},  {\em JHEP} {\bf 11} (2017) 188,
  [\href{http://arxiv.org/abs/1709.10184}{{\tt arXiv:1709.10184}}].

\bibitem{Hernandez:2020nem}
J.~Hernandez, R.~C. Myers, and S.-M. Ruan, {\it {Quantum Extremal Islands Made
  Easy, PartIII: Complexity on the Brane}},
  \href{http://arxiv.org/abs/2010.16398}{{\tt arXiv:2010.16398}}.

\bibitem{Davis:2002gn}
S.~C. Davis, {\it {Generalized Israel junction conditions for a Gauss-Bonnet
  brane world}},  {\em Phys. Rev. D} {\bf 67} (2003) 024030,
  [\href{http://arxiv.org/abs/hep-th/0208205}{{\tt hep-th/0208205}}].

\bibitem{Wall:2012uf}
A.~C. Wall, {\it {Maximin Surfaces, and the Strong Subadditivity of the
  Covariant Holographic Entanglement Entropy}},  {\em Class. Quant. Grav.} {\bf
  31} (2014), no.~22 225007, [\href{http://arxiv.org/abs/1211.3494}{{\tt
  arXiv:1211.3494}}].

\bibitem{Czech:2012bh}
B.~Czech, J.~L. Karczmarek, F.~Nogueira, and M.~Van~Raamsdonk, {\it {The
  Gravity Dual of a Density Matrix}},  {\em Class. Quant. Grav.} {\bf 29}
  (2012) 155009, [\href{http://arxiv.org/abs/1204.1330}{{\tt
  arXiv:1204.1330}}].

\bibitem{Akers:2016ugt}
C.~Akers, J.~Koeller, S.~Leichenauer, and A.~Levine, {\it {Geometric
  Constraints from Subregion Duality Beyond the Classical Regime}},
  \href{http://arxiv.org/abs/1610.08968}{{\tt arXiv:1610.08968}}.

\bibitem{Camps:2019opl}
J.~Camps, {\it {The Parts of the Gravitational Field}},
  \href{http://arxiv.org/abs/1905.10121}{{\tt arXiv:1905.10121}}.

\bibitem{Engelhardt:2013tra}
N.~Engelhardt and A.~C. Wall, {\it {Extremal Surface Barriers}},  {\em JHEP}
  {\bf 03} (2014) 068, [\href{http://arxiv.org/abs/1312.3699}{{\tt
  arXiv:1312.3699}}].

\end{thebibliography}\endgroup
\bibliographystyle{JHEP}

\end{document}